\pdfoutput=1
\documentclass[12pt]{article}
\usepackage[utf8]{inputenc}
\usepackage[a4paper,top=1in,bottom=1in,left=1in,right=1in]{geometry}
\usepackage{fullpage,caption,subcaption}
\usepackage{graphicx}
\usepackage{pdfpages}
\usepackage{xcolor}
\usepackage{tikz}
\usetikzlibrary{automata, positioning, arrows}

\usepackage{wrapfig}
\usepackage{pgfplots}
\pgfplotsset{compat=1.8}
\pgfplotsset{axis line style={-}}
\usepgfplotslibrary{statistics}
\usepackage{filecontents}

\usepackage{url}

\usepackage[hidelinks]{hyperref}
\hypersetup{colorlinks=false,breaklinks=true}

\usepackage{setspace}
\definecolor{nodebg}{HTML}{F7F7F7}
\tikzset{
->,
node distance=3.8cm,
every state/.style={thick, fill=nodebg},
initial text=,
}

\usepackage{biblatex}
\begin{document}

\includepdf{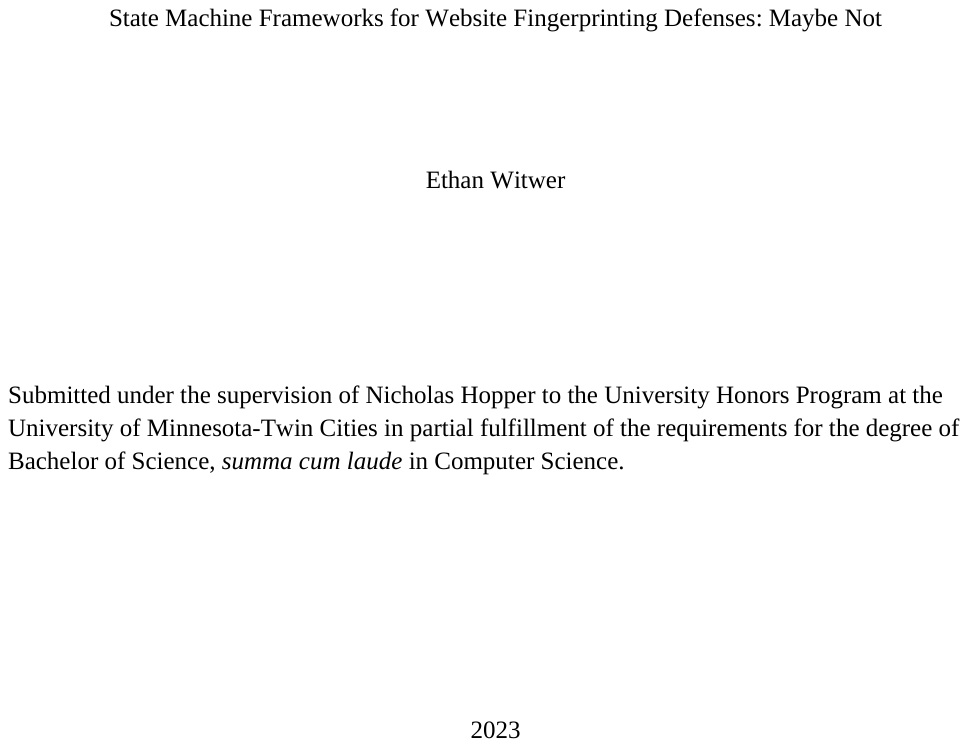}
\addtocounter{page}{-1}

\section*{Abstract}
Tor is an anonymity network used by millions of people every day to evade censorship and protect their browsing activity from privacy threats such as mass surveillance. Tor conceals communication metadata, which indicates who is sending messages to whom, or which websites a user is visiting. Unfortunately, though, Tor has been shown to be vulnerable to website fingerprinting attacks, in which an adversary observes the connection between a user and the Tor network and leverages features of the encrypted traffic, such as the timing and volume of packets, to identify the websites that are being visited. This undermines the protection goals of Tor and puts its users at risk of exposure.

In response, researchers have proposed a number of defenses against website fingerprinting attacks, and a ``circuit padding framework'' has been added to the Tor software which supports the implementation of defenses. However, many proposed defenses cannot be implemented with this framework, because it requires defenses to be modeled as state machines which probabilistically send padding packets through a Tor circuit. Since padding packets look like normal packets to an observer but contain no useful data, they can be used to conceal traffic patterns that are taken advantage of by attacks; but a large number of defenses are deterministic, have complex behavior, or involve delaying packets, which is not supported by the circuit padding framework. Additionally, no padding-only defenses have been shown to be effective enough to merit the overhead that they incur, so none are currently implemented in Tor.

As Arti, a reimplementation of Tor in the Rust programming language, is being developed, the issue arises of whether a new state machine framework should be included or if alternative models should instead be considered for future defense implementation. We address this question by using an improved Rust-based state machine framework, Maybenot, to implement three state-of-the-art website fingerprinting defenses: FRONT, RegulaTor, and Surakav. By evaluating our implementations in terms of their similarity to the simulated versions of these defenses, overhead, and protection against attacks, we demonstrate the potential of state machine frameworks to support effective defenses, and we highlight important features that they should contain to do so. However, our evaluation also raises uncertainty about the long-term feasibility of state machine frameworks for defense implementation. We recommend enhancements to Maybenot and substantial further evaluation, along with consideration of alternative designs, before any decision is made regarding a mechanism for implementing website fingerprinting defenses in Arti.

\newpage
\tableofcontents

\newpage
\section{Introduction}
In the face of steadily increasing mass surveillance and threats to online privacy, many people are turning to privacy-enhancing technologies such as Tor to protect their browsing activity. Tor is an anonymity network that aims to hide communication metadata, which reveals who is communicating with whom, or which websites a user is visiting~\cite{dingledine2004tor}. Tor serves millions of users every day, including journalists, activists, whistleblowers, and ordinary people with an interest in protecting their privacy~\cite{tor-metrics, tor-users}.

Tor works by routing connections through a \textit{circuit} consisting of three relays\textemdash guard, middle, and exit\textemdash using layered encryption so that each relay only sees the IP addresses of the previous and next hops in the circuit. The guard relay forwards traffic from the client to the middle relay; the middle relay forwards from the guard to the exit relay; and the exit relay forwards from the middle relay to the destination. A layer of encryption is stripped at each relay, revealing information about the next hop in the circuit. In this way, no single entity is aware of the identities of both the client and destination.

\begin{figure}[h!]
  \centering
  \includegraphics[width=\linewidth]{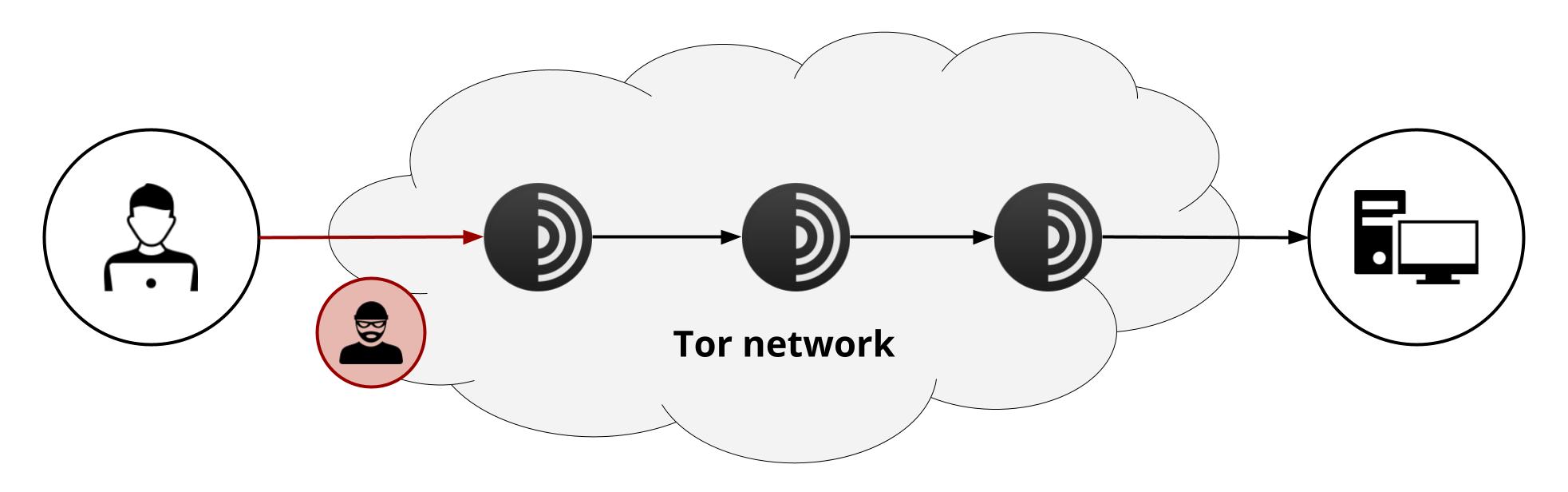}
  \caption{The website fingerprinting threat model in Tor}
  \label{fig:wf_threat_model}
\end{figure}

However, it has been demonstrated that Tor is vulnerable to a type of traffic analysis attack called \textit{website fingerprinting}, in which a local, passive adversary identifies which websites a user is visiting by observing the traffic sent on the connection between the client and guard relay (Figure~\ref{fig:wf_threat_model}). In certain settings, website fingerprinting attacks have proved highly effective, identifying monitored web pages with greater than 98\% accuracy~\cite{rahman2019tik,sirinam2018deep}. This defeats the goals of Tor and poses a significant threat to the privacy of its users, making the need for effective defense mechanisms critical.

Tor's main codebase is written in C, and it contains a ``circuit padding framework'' that allows for the implementation of website fingerprinting defenses~\cite{tor-circpad}. Defenses are modeled as state machines that control the probabilistic injection of padding packets onto the connection between a Tor client and relay, with the goal of concealing traffic patterns used in website fingerprinting attacks. However, many effective defenses that have been proposed by researchers cannot be implemented with this framework, notably those that involve delaying packets, which is not supported.

The Tor Project is currently developing Arti, a Rust implementation of Tor that will eventually completely replace the C code~\cite{announcing-arti}, and there are not yet any concrete plans for a website fingerprinting defense framework~\cite{arti-framework}. This raises the question of whether a state machine framework similar to the circuit padding framework should be included in Arti or if alternative designs should be explored. In this work, we address this issue by assessing the capability of state machine frameworks to support current state-of-the-art website fingerprinting defenses.

We consider a recently developed Rust-based framework for traffic analysis defenses called Maybenot~\cite{pulls2023maybenot}. Maybenot is a generalization of Tor's circuit padding framework and introduces new features including packet delays, making it more representative of the maximal capabilities of state machine frameworks. Thus, in working with Maybenot, we gather important insights into the potential of state machine frameworks in general to support proposed defenses.

We use Maybenot to implement three state-of-the-art defenses: FRONT~\cite{gong2020zero}, RegulaTor~\cite{holland2022regulator}, and Surakav~\cite{gong2022surakav}. We compare our implementations to the simulated versions of these defenses and measure their overhead and protection against attacks. This evaluation demonstrates that state machine frameworks have the potential to support effective defenses, but it also brings up important issues and considerations related to their long-term viability for defense implementation, which are discussed in detail.

Overall, we offer the following main contributions:

\begin{itemize}
    \item We provide two implementations of FRONT, one of RegulaTor, and a precursor to Surakav in the Rust-based state machine framework Maybenot. Their source code has been made publicly available on GitHub.
    \item We demonstrate that state machine frameworks can be used to implement effective website fingerprinting defenses, but we observe that certain features are needed to do so optimally and suggest improvements to the Maybenot framework.
    \item We discuss the implications of adopting a state machine framework for website fingerprinting defenses in Arti and recommend further evaluation and consideration of alternative designs before the decision to include one (or not) is made.
\end{itemize}

\newpage
\section{Background}
\paragraph{Tor and Arti.} Tor is an anonymity network that is designed to protect the metadata of its users' communications~\cite{dingledine2004tor}. It is intended to ensure that no single entity can determine both the source and destination of any correspondence, meaning that the websites a user visits should not be discernible.

To achieve this, a client builds a \textit{circuit} consisting of three intermediate \textit{relays}\textemdash guard, middle, and exit\textemdash and establishes separate symmetric encryption keys with each one in a telescopic fashion. Data is sent through the circuit in 512-byte \textit{cells} that are encrypted in layers, once with every key: each relay can only decrypt one layer, revealing the next relay to send the cell to (or the final destination, in the case of the exit relay).

The original Tor software was written in C, and it is still widely used as of 2023. However, in 2020, the Tor Project began working on Arti, a reimplementation of Tor in Rust~\cite{announcing-arti}. Arti was intended to avoid common bugs in C programs and improve on the design of C Tor. It was deemed suitable for production use in 2022; though it still lacks some of the features available in C Tor, its eventual goal is to replace it entirely~\cite{arti-ready}.

\paragraph{Website Fingerprinting.} In website fingerprinting (WF) attacks, an adversary records the sequence of packets sent over the connection between a Tor client and guard relay, producing a \textit{trace} of the traffic flow. Although the contents of packets and their ultimate destination are encrypted, the adversary can still deduce the websites a user has visited with a classifier which makes use of features such as the timing and volume of packets. These features form a unique \textit{fingerprint} that is largely consistent for any given web page, making it possible to identify the page from a trace.

WF attacks are considered in the context of a local, passive adversary. The adversary is deemed \textit{local} since he must only observe the connection between the client and guard relay. This can be done by a number of actors, including another user on the local network, a user's Internet service provider, or the guard relay itself. Moreover, the adversary is \textit{passive} since he does not modify the traffic flow; he only needs the ability to observe it. This scenario falls within Tor's threat model and represents a serious threat to the privacy of its users~\cite{dingledine2004tor}.

Evaluation of WF attacks is typically done in either the \textit{closed-world} setting, in which an adversary has access to sample traces for all of the web pages that a user might visit; or the more realistic \textit{open-world} setting, in which a user can additionally visit pages that are not known by the adversary. Unfortunately, it has been demonstrated that WF attacks are highly effective against Tor in both settings~\cite{bhat2019var, hayes2016k, panchenko2016website, rahman2019tik, sirinam2018deep}. Other settings are even more favorable to attackers, such as the one-page setting proposed by Wang, in which an adversary attempts to detect visits to only a single web page~\cite{wang2021one}.

Many effective WF attacks based on machine learning classifiers have been proposed. Panchenko et al. proposed CUMUL in 2016, which uses the ``cumulative representation'' of a trace along with an SVM classifier to achieve 92\% accuracy in the closed-world setting~\cite{panchenko2016website}. In 2018, Sirinam et al. proposed Deep Fingerprinting (DF), an attack that uses deep learning to achieve 98\% accuracy using only sequences of packet directions~\cite{sirinam2018deep}. DF was extended by Rahman et al. in 2019 to include timing features, resulting in Tik-Tok~\cite{rahman2019tik}. Other classifiers have also obtained high accuracy against Tor~\cite{bhat2019var, hayes2016k, herrmann2009website}.

Furthermore, although some of the assumptions made in attack evaluations have been criticized for being unrealistic~\cite{juarez2014critical}, some work has demonstrated that certain assumptions are not required for WF attacks to be practical, such as~\cite{wang2016realistically}; and highly effective attacks have been carried out in the real world with a small set of monitored pages~\cite{cherubin2022online}.

\paragraph{Defenses.} A number of defenses against WF attacks have been proposed, most of which attempt to directly modify a trace in real time through a combination of sending padding packets and delaying packets. Some of these defenses are padding-only, meaning that they do not induce delay; they introduce extra packets into traces in ways that are intended to mask specific features used in attacks. WTF-PAD attempts to hide delays between packets that are unusually long (``statistically unlikely'')~\cite{juarez2016toward}, and another defense, FRONT, inserts padding near the beginning of a download based on the intuition that the most useful features are present in the initial portion of download traffic~\cite{gong2020zero}. Padding-only defenses are widely favored due to the assumption that they do not degrade user experience as much as those involving delay, but this is not true in practice~\cite{witwer2022padding}.

Another class of defenses include delay as a crucial part of their design, and many of these are intended to ``regularize'' traffic traces, making them appear similar to all other traces or to others in an anonymity set. An extreme example of this is BuFLO, proposed in 2012, which sends traffic at a constant rate throughout an entire download, sending padding and delaying packets as necessary to do so~\cite{dyer2014peek}. In 2015, Wang presented an improved version of BuFLO called Tamaraw, which includes separate constant rates for download and upload traffic and pads the total length of a download up to a multiple of some chosen parameter~\cite{wang2015website}. This provides gains in overhead and protection against attacks, but BuFLO and Tamaraw are both impractical defenses due to their very high overhead. A newer regularizing defense, RegulaTor, sends traffic at an initial constant rate that decreases according to a decay function, achieving a high level of protection against attacks with more practical overhead requirements~\cite{holland2022regulator}.

Some defenses shape traces in an attempt to make them look like other \textit{specific} traces. Decoy, proposed in 2011, loads a second page in the background to confuse an attacker instead of directly padding or delaying packets~\cite{panchenko2011website}. In 2017, Wang and Goldberg described Walkie-Talkie, which makes pairs of web pages have the same trace by modifying the browser to communicate in a half-duplex mode and performing ``burst molding''~\cite{wang2017walkie}. Larger anonymity sets can also be used: Glove, proposed by Nithyanand et al. in 2014, morphs all traces in a computed ``cluster'' into a single ``supertrace,'' rendering them mutually indistinguishable~\cite{nithyanand2014glove}; a similar approach is taken by Supersequence, also presented in 2014~\cite{wang2014effective}. These defenses are limited by the fact that they require prior knowledge of web page traces, and \textit{which} pages are grouped together in an anonymity set can have a significant impact on their efficacy~\cite{rahman2019tik, wang2021one}; a recent defense, Surakav, avoids the latter issue by employing adversarial machine learning techniques to generate fake traces and shaping traffic to make it appear similar to these traces~\cite{gong2022surakav}.

The defenses summarized so far all operate at the network layer, directly modifying traces; still more defenses work at the application layer and take advantage of the specifics of the protocol in use to defend against attacks. For instance, HTTPOS manipulates the TCP window and downloads web objects in separate chunks via the HTTP Range header and HTTP pipelining to confuse attacks~\cite{luo2011httpos}; however, it has been shown to be ineffective~\cite{cai2012touching}. Other network-layer techniques are also possible, such as splitting traffic over multiple Tor circuits or network links, approaches which are taken by TrafficSliver~\cite{cadena2020trafficsliver} and HyWF~\cite{henri2020protecting}, respectively. In this work, we only consider standard network-layer defenses, as they provide broader protection and are suitable for direct implementation in Tor. We implement FRONT, RegulaTor, and Surakav as exemplars of the three categories of network-layer defenses that have been discussed.

\paragraph{Defense Frameworks.} Shmatikov and Wang proposed \textit{adaptive padding} in 2006 to defend against timing analysis attacks in mix networks~\cite{shmatikov2006timing}. This technique involves maintaining a likely distribution of inter-packet time intervals. When a packet is received by a mix, an interval is sampled: if another packet arrives before the interval expires, it is forwarded and a new interval is sampled; otherwise, a padding packet is sent before sampling another interval. This has the effect of making inter-packet delays roughly follow an expected distribution. Adaptive padding can also operate in a dual mode, which minimizes padding sent within bursts of traffic and focuses on protecting gaps between bursts. In this mode, when a packet is received by a mix, a larger interval is sampled; if an interval expires and padding is sent, a shorter one is sampled next.

The dual mode of adaptive padding (and the basic algorithm) is implemented using histograms: when the first packet of a connection is received by a mix, a histogram is constructed with bins representing ranges of inter-packet delays. Each bin is filled with \textit{tokens}: a configurable number of values are sampled from the inter-packet interval distribution, and if a value falls within a bin's range, a token is added to that bin. Then, when a packet is received by the mix, a random token is selected and removed from its bin, and an inter-packet interval is sampled from the bin's range. If the interval expires without any traffic being received, a padding packet is sent; otherwise, the received packet is forwarded, the removed token is returned to its bin, and a token is removed from the bin corresponding to the observed inter-packet delay. In the former case, the next interval is sampled from the ``low-bins set;'' otherwise, one is sampled from the ``high-bins set.''

In 2016, Juarez et al. extended the dual mode of adaptive padding to develop the WF defense WTF-PAD~\cite{juarez2016toward}. This included (1) adding control messages so that the client is in charge of padding decisions and (2) allowing padding to be sent in response to packets received \textit{and} sent, since Tor clients do not forward traffic as do mixes; separate distributions are maintained for this purpose. A circuit padding framework based on WTF-PAD was developed for C Tor in 2019~\cite{tor-circpad}. Clients can negotiate the use of padding state machines with any relay in a circuit, and padding will be sent probabilistically according to the state machine in use; histograms can be used along with parameterized probability distributions, as was done by Pulls in APE, a defense similar to WTF-PAD~\cite{ape}. However, packet delays are not supported nor are particularly complex behaviors.

Besides APE and WTF-PAD, which has been defeated~\cite{sirinam2018deep}, a few other WF defenses have been developed for the circuit padding framework. In 2018, Mathews et al. presented the Random Extend Bursts defense, which adds padding to \textit{bursts} of cells (uninterrupted sequences of cells sent in a single direction on a connection), but it is not effective against DF~\cite{mathews2018understanding}. Pulls used genetic algorithms with the circuit padding framework in 2020 and manually modified the best evolved state machine to develop the defenses Spring and Interspace, which are effective against DF but have prohibitive bandwidth overhead~\cite{pulls2020effective}. No WF defenses are currently deployed in Tor.

\paragraph{Maybenot.} In this work, we use Maybenot~\cite{pulls2023maybenot}, a more recent framework written in Rust. Maybenot is a generalization of Tor’s circuit padding framework; it allows for more sophisticated defenses, eliminates histograms in favor of parameterized probability distributions, and includes support for packet delays (temporarily blocking traffic from being sent). An application using Maybenot provides as input \textit{events} related to an encrypted communication channel, such as receiving or sending a packet, and is presented with \textit{actions} that should be taken to defend the channel, i.e., sending padding or blocking outgoing traffic. \textit{Machines} define which action to take when a given event occurs. Several machines may operate in parallel to build more complex defenses.

Each machine consists of a number of \textit{states}, which are characterized by an action, three distributions, and vectors of probabilities for state transitions. The action is either to send padding or block outgoing traffic. An action distribution specifies either the amount of padding or the duration of blocking; a timeout distribution is used to sample the time that should pass before the action is applied; and a limit distribution indicates how many self-transitions can occur before a \textit{LimitReached} event is triggered. A map is maintained between each event and a corresponding state transition vector, allowing different events to have their own probabilities of transitioning to any given state. More specific features of Maybenot will be discussed as relevant throughout the paper.

\newpage
\section{Implementing FRONT}

\subsection{Description}

The FRONT defense is based on two primary observations: (1) the beginning of a trace contains most of the features used in WF attacks, and (2) trace-to-trace randomness can be employed to create strong defenses, as it reduces an attacker's ability to effectively train a classifier on defended traces~\cite{gong2020zero}. Thus, a high volume of padding cells are sent at the beginning of each trace, and the number and timing of these cells differ among traces.

This is accomplished by using a Rayleigh distribution to schedule padding cells before a download begins. A number of time values are sampled, and a padding cell is then sent at each of these times relative to download start. Trace-to-trace randomness is achieved by varying the distribution's scale parameter and the number of values sampled among downloads. The entire sequence of steps involved in FRONT is as follows:

\begin{enumerate}
  \item A padding count is sampled from a discrete uniform distribution: \begin{math}n_c\end{math} is sampled from the range \begin{math}[1, N_c]\end{math} by the client, and \begin{math}n_s\end{math} is sampled from \begin{math}[1, N_s]\end{math} by the relay. The parameters \begin{math}N_c\end{math} and \begin{math}N_s\end{math} are the client and relay padding budgets, respectively.
  \item A padding window, which is the scale parameter \begin{math}\sigma\end{math} of the Rayleigh distribution, is sampled from a continuous uniform distribution: \begin{math}w_c\end{math} is sampled from \begin{math}[W_{min}, W_{max}]\end{math} by the client, and \begin{math}w_s\end{math} is sampled from the same range by the relay.
  \item Padding cells are scheduled: \begin{math}n_c\end{math} values are sampled from the Rayleigh distribution by the client, and \begin{math}n_s\end{math} values are sampled by the relay.
  \item During the download, a padding cell will be sent at each sampled time. No further padding is sent after the download completes.
\end{enumerate}

\subsection {First Machine: Maybenot FRONT}
We sought to implement FRONT in Maybenot using a single machine design, since both client and relay perform the same straightforward sequence of steps. Because padding cells are scheduled \textit{before} a download starts in FRONT, this could not be done directly; instead, we aimed to approximate the sending rate of padding cells that would result from sampling time values from a Rayleigh distribution.

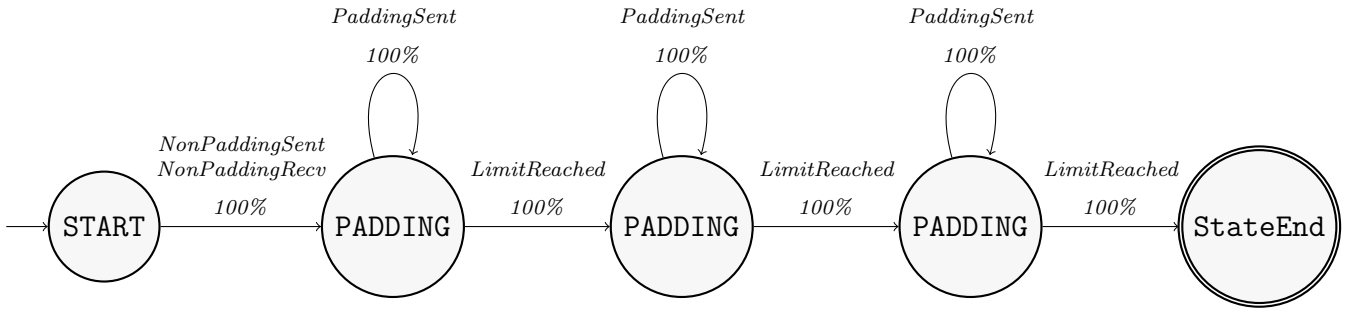
\begin{figure}[h!]
    \centering
    \makebox[\textwidth][c]{
    \begin{tikzpicture}
        \node[state, initial] (q0) {\texttt{START}};
        \node[state, right of=q0] (q1) {\texttt{PADDING}};
        \node[state, right of=q1] (q2) {\texttt{PADDING}};
        \node[state, right of=q2] (q3) {\texttt{PADDING}};
        \node[state, accepting, right of=q3] (q4) {\texttt{StateEnd}};
        \draw (q0) edge[above, text width=3cm, align=center] node{\scriptsize{\textit{NonPaddingSent\\NonPaddingRecv\\100\%}}} (q1)
              (q1) edge[loop above, text width=3cm, align=center] node{\scriptsize{\textit{PaddingSent\\100\%}}} (q1)
              (q1) edge[above, text width=3cm, align=center] node{\scriptsize{\textit{LimitReached\\100\%}}} (q2)
              (q2) edge[loop above, text width=3cm, align=center] node{\scriptsize{\textit{PaddingSent\\100\%}}} (q2)
              (q2) edge[above, text width=3cm, align=center] node{\scriptsize{\textit{LimitReached\\100\%}}} (q3)
              (q3) edge[loop above, text width=3cm, align=center] node{\scriptsize{\textit{PaddingSent\\100\%}}} (q3)
              (q3) edge[above, text width=3cm, align=center] node{\scriptsize{\textit{LimitReached\\100\%}}} (q4);
    \end{tikzpicture}
    }
    \caption{Maybenot FRONT machine with three \texttt{PADDING} states}
    \label{fig:front_imp}
\end{figure}

Our first FRONT machine design consists of a \texttt{START} state, a number of \texttt{PADDING} states, and the pseudo-state \texttt{StateEnd} provided by the Maybenot framework, as shown in Figure~\ref{fig:front_imp}. Each machine is characterized by its padding budget $N$, maximum padding window $W_{max}$, and number of \texttt{PADDING} states $\psi$. We call this design Maybenot FRONT.

The Maybenot FRONT machine transitions from \texttt{START} to the first \texttt{PADDING} state when a download begins, which is considered to occur when the first non-padding cell is sent or received (when a \textit{NonPaddingSent} or \textit{NonPaddingRecv} event is triggered). It then proceeds sequentially through the remaining \texttt{PADDING} states until it reaches \texttt{StateEnd}.

As their name suggests, \texttt{PADDING} states generate \textit{Padding} actions; a uniform action distribution with parameters $a=512$ and $b=512$ is used so that a single padding cell is sent per action. The delay before a cell should be sent is sampled from a normal timeout distribution, and trace-to-trace randomness is achieved with a uniform limit distribution.

When a \texttt{PADDING} state is transitioned to, it generates a \textit{Padding} action with a timeout value sampled from its timeout distribution. The application using Maybenot (i.e., Arti) will send a padding cell after this timeout expires and trigger a \textit{PaddingSent} event, causing a self-transition. This will occur repeatedly until the state's limit is reached, at which time a \textit{LimitReached} event will be triggered, causing a transition to the next state.

Each \texttt{PADDING} state is modeled as corresponding to a fixed time slice of a download; its timeout distribution parameters are selected to approximate the distribution of inter-packet delays that would result during that interval if time values were sampled from a Rayleigh distribution with $\sigma = W_{max}$. In a machine with $\psi$ \texttt{PADDING} states, the timeout distribution parameters of a \texttt{PADDING} state that spans the interval from $t_1$ to $t_2$ are:

\begin{equation}
    \mu = \frac{ \psi }{ N } \cdot (t_2 - t_1)
\end{equation}

\begin{equation}
    \sigma = \frac{ W_{max}^2 }{ \sqrt{ \pi } } \cdot (\frac{ N }{ \psi } \cdot \frac{ t_1 + t_2 }{ 2 })^{-1}
\end{equation}
\\

$\mu$ is selected to be the inter-packet delay that would result in exactly $N / \psi$ cells being sent during the interval from $t_1$ to $t_2$. The equation for $\sigma$ is partially derived from the results of preliminary simulations and trace comparisons; it allows for greater variation of inter-packet delays near the beginning of a download, and variation is increased for larger values of $W_{max}$. To prevent excessive variation, timeout values are bounded to be in the range $[0, 2 \cdot \mu]$ by specifying a \texttt{max} parameter for the timeout distribution.

If each state had a constant limit corresponding to the number of cells that would most likely be sent during its interval, this would allow for a precise approximation of the sending rate of padding cells that would result from a Rayleigh distribution. However, such a design would not account for the trace-to-trace randomness FRONT is intended to achieve: a machine's padding count would be constant, and variation of the padding window would be small and only due to differences in sampled timeout values.

To mimic the sampling performed by FRONT, we instead use a uniform distribution with range $[1, N / \psi]$ for each state's limit. Thus, the padding count for a download is effectively sampled from a uniform \textit{sum} distribution with range \begin{math}[\psi, N]\end{math}. Note that this change allows for variation in the padding count as well as the padding window, as the times at which state transitions occur become more variable.

\subsection{Second Machine: Pipelined FRONT}

\begin{figure}[h!]
    \centering
    \makebox[\textwidth][c]{
    \begin{tikzpicture}
        \node[state, initial] (q0) {\texttt{START}};
        \node[state, right of=q0, yshift=38] (q1) {\texttt{PADDING}};
        \node[state, right of=q0, yshift=-38] (q7) {\texttt{PADDING}};
        \node[state, right of=q1] (q2) {\texttt{PADDING}};
        \node[state, right of=q2] (q3) {\texttt{PADDING}};
        \node[state, accepting, right of=q3, yshift=-38] (q4) {\texttt{StateEnd}};
        \node[state, right of=q7] (q5) {\texttt{PADDING}};
        \node[state, right of=q5] (q6) {\texttt{PADDING}};
        \draw (q1) edge[loop above, text width=3cm, align=center] node{\scriptsize{\textit{PaddingSent\\100\%}}} (q1)
              (q0) edge[above, text width=3cm, align=center] node{\scriptsize{\textit{\hspace{-2em}NonPaddingSent\\\hspace{-2em}NonPaddingRecv\\\hspace{-2em}50\%}}} (q1)
              (q0) edge[above, text width=3cm, align=center] node{ } (q7)
              (q1) edge[above, text width=3cm, align=center] node{\scriptsize{\textit{LimitReached\\100\%}}} (q2)
              (q7) edge[below, text width=3cm, align=center] node{\scriptsize{\textit{\\LimitReached\\100\%}}} (q5)
              (q2) edge[loop above, text width=3cm, align=center] node{\scriptsize{\textit{PaddingSent\\100\%}}} (q2)
              (q5) edge[loop below, text width=3cm, align=center] node{\scriptsize{\textit{PaddingSent\\100\%}}} (q5)
              (q7) edge[loop below, text width=3cm, align=center] node{\scriptsize{\textit{PaddingSent\\100\%}}} (q7)
              (q2) edge[above, text width=3cm, align=center] node{\scriptsize{\textit{LimitReached\\100\%}}} (q3)
              (q5) edge[below, text width=3cm, align=center] node{\scriptsize{\textit{\\LimitReached\\100\%}}} (q6)
              (q3) edge[loop above, text width=3cm, align=center] node{\scriptsize{\textit{PaddingSent\\100\%}}} (q3)
              (q6) edge[loop below, text width=3cm, align=center] node{\scriptsize{\textit{PaddingSent\\100\%}}} (q6)
              (q3) edge[above, text width=3cm, align=center] node{\scriptsize{\textit{\hspace{2em}LimitReached\\\hspace{2em}100\%}}} (q4)
              (q6) edge[below, text width=3cm, align=center] node{ } (q4);
    \end{tikzpicture}
    }
    \caption{Pipelined FRONT machine with two pipelines, three \texttt{PADDING} states each}
    \label{fig:pipelined_front_imp}
\end{figure}
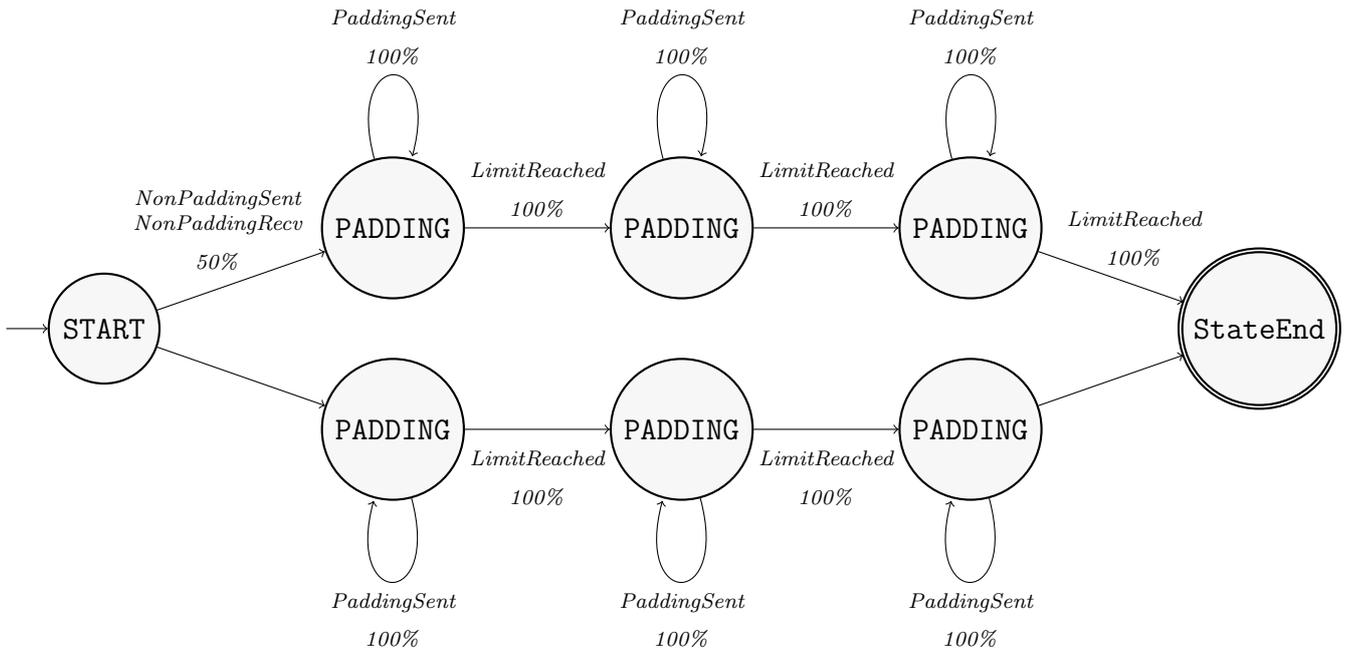

A limitation of Maybenot FRONT is that the timeout distribution parameters of \texttt{PADDING} states are calculated using values that are fixed for each machine. Although the padding count and window do vary among downloads, inter-packet timing is less variable, which reduces the efficacy of the defense. To remedy this, we introduce Pipelined FRONT, a machine based on the same principles as Maybenot FRONT but with multiple \textit{pipelines} that have different padding budgets, as illustrated in Figure~\ref{fig:pipelined_front_imp}.

In this machine, the first state to transition to is chosen from a set of \texttt{PADDING} states which all have equal probability, and each one leads to a different pipeline. This allows for variation of the padding count and window, as with Maybenot FRONT, as well as inter-packet timing, which greatly improves trace-to-trace randomness.

\subsection{Evaluation}

\subsubsection{Experimental Setup}

In our evaluations, we used the BigEnough dataset collected by Mathews et al. between November 2021 and January 2022~\cite{mathews2022sok}. Specifically, we used the monitored set, which consists of 19,000 traces collected from 95 websites. To build the monitored set, 10 subpages from every website were visited 20 times each, for a total of 200 traces per site. This was done using the ``Safest'' configuration in the Tor Browser.

\begin{table}[b!]
    \centering
    \begin{tabular}{ |c||c|c|c|c| }
        \hline
        {Defense} & \multicolumn{4}{c|}{Parameters}\\
        \cline{2-5}
        & \textbf{\begin{math}N\end{math}} & \textbf{\begin{math}W_{min}\end{math}} & \textbf{\begin{math}W_{max}\end{math}} & \textbf{\begin{math}\psi\end{math}} \\
        \hline\hline
        \textbf{Maybenot FT-1} & 1500 & 1 s & 14 s & 30 \\
        \hline
        \textbf{Pipelined FT-1} & 3000 & 1 s & 14 s & \begin{math}30 \times 30\end{math} \\
        \hline
        \textbf{Simulated FT-1} & 1700 & 1 s & 14 s & --- \\
        \hline\hline
        \textbf{Maybenot FT-2} & 2500 & 1 s & 14 s & 50 \\
        \hline
        \textbf{Pipelined FT-2} & 4500 & 1 s & 14 s & \begin{math}45 \times 45\end{math} \\
        \hline
        \textbf{Simulated FT-2} & 2500 & 1 s & 14 s & --- \\
        \hline
    \end{tabular}
    \caption{FRONT parameters}
    \label{tab:front-params}
\end{table}

We used the simulation scripts provided by Gong et al.~\cite{gong2020zero} to produce FRONT-defended traces and the Maybenot simulator~\cite{pulls2023maybenot} to defend traces with Maybenot FRONT and Pipelined FRONT. A delay of 10 ms was simulated between the client and server by the Maybenot simulator; the simulator uses delays to set up event queues, and the value selected had a marginal impact on the resultant defended traces.

Using the FRONT simulation scripts, we generated two defended datasets, one for each of the two configurations of FRONT presented by Gong et al.: \texttt{FT-1}, a lightweight configuration with $N_s = N_c = 1700$, $W_{min} = 1 s$, and $W_{max} = 14 s$; and \texttt{FT-2}, with $N_s = N_c = 2500$ and the same window parameters as \texttt{FT-1}~\cite{gong2020zero}.

We produced four defended datasets with the Maybenot simulator: Maybenot \texttt{FT-1}, Maybenot \texttt{FT-2}, Pipelined \texttt{FT-1}, and Pipelined \texttt{FT-2}. The Maybenot simulator accepts machines in their serialized form (character strings) and defends input traces with them; to obtain Maybenot FRONT and Pipelined FRONT machines, we developed two simple Rust programs that generate them based on supplied parameters.

We chose parameters for our implementations to match the bandwidth overhead incurred by simulated FRONT; the final parameters selected for each implementation are summarized in Table~\ref{tab:front-params}. Note that, for Pipelined FRONT, $\psi$ represents number of pipelines followed by the number of \texttt{PADDING} states per pipeline.

After running the Maybenot simulator, we removed trailing padding cells from each trace in the four defended datasets to better compare with simulated FRONT. Our implementations do not include any mechanism to detect the end of a download, and Maybenot provides no such event; this issue is discussed in Section 6.2.

\subsubsection{Trace Comparison}

We determined the similarity of corresponding traces defended with simulated FRONT and our machines, adopting the methodology of Smith et al.~\cite{smith2022qcsd} We represented each trace with two \textit{aggregated time series}, one for upload traffic and another for download traffic. We computed these by partitioning total download time into fixed-length windows of \begin{math}I\end{math} ms and creating sequences consisting of the total number of bytes sent and received by the client during each window. We set \begin{math}I = \{25, 50\}\end{math} for our evaluations.

\begin{figure*}[h!]
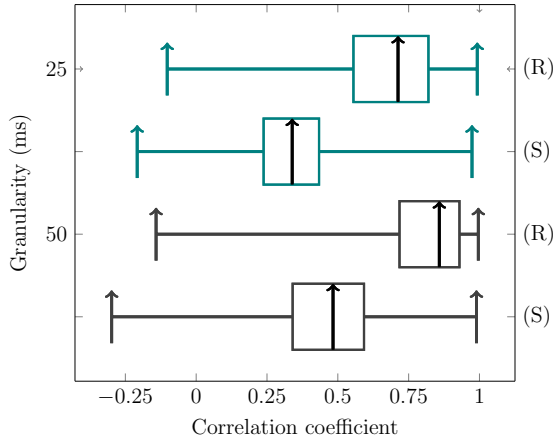
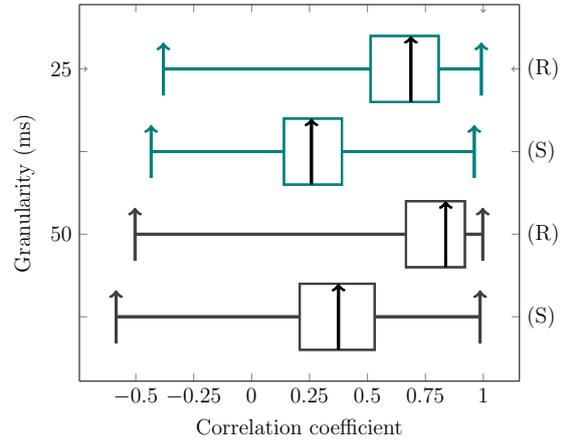

    \centering
    \makebox[\textwidth][c]{
    \begin{subfigure}[b!]{0.5\textwidth}
        \include{compare_corr_ft1}
        \caption{\texttt{FT-1} configuration}
    \end{subfigure}
    \hfill
    \begin{subfigure}[b!]{0.5\textwidth}
        \include{compare_corr_ft2}
        \caption{\texttt{FT-2} configuration}
    \end{subfigure}
    }
    \caption{Correlation coefficient, simulated FRONT and Maybenot FRONT}
    \label{tab:front-correlation}
\end{figure*}

\begin{figure*}[h!]
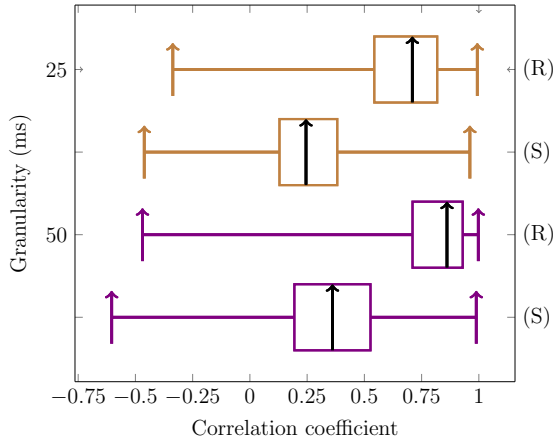
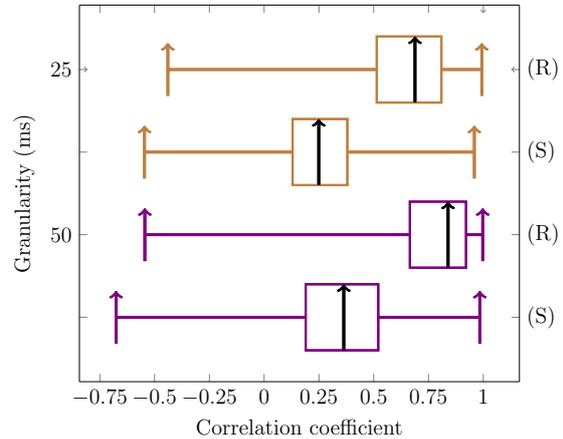

    \centering
    \makebox[\textwidth][c]{
    \begin{subfigure}[b!]{0.5\textwidth}
        \include{compare_corr_ft1_pipe}
        \caption{\texttt{FT-1} configuration}
    \end{subfigure}
    \hfill
    \begin{subfigure}[b!]{0.5\textwidth}
        \include{compare_corr_ft2_pipe}
        \caption{\texttt{FT-2} configuration}
    \end{subfigure}
    }
    \caption{Correlation coefficient, simulated FRONT and Pipelined FRONT}
    \label{tab:front-pipe-correlation}
\end{figure*}

\begin{figure*}[b!]
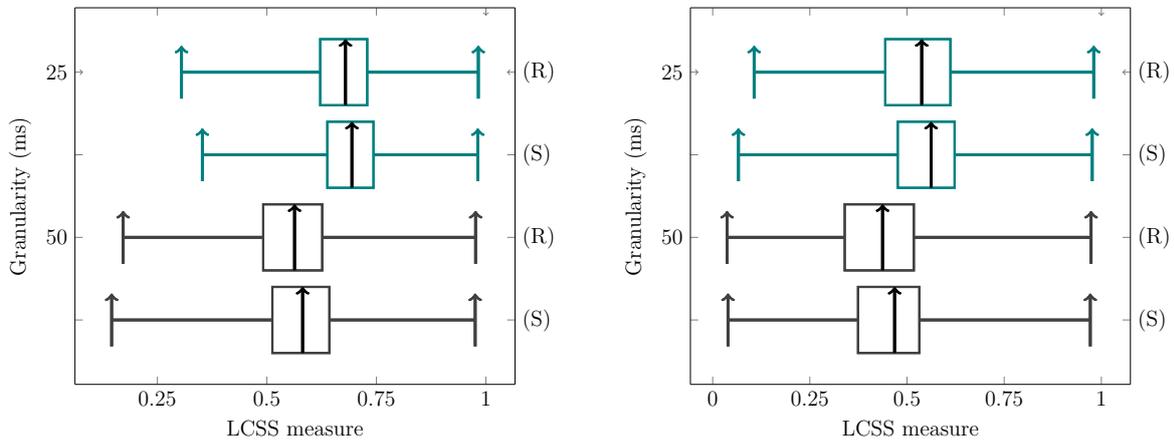

    \centering
    \makebox[\textwidth][c]{
    \begin{subfigure}[b!]{0.5\textwidth}
        \include{compare_lcss_ft1}
        \caption{\texttt{FT-1} configuration}
    \end{subfigure}
    \hfill
    \begin{subfigure}[b!]{0.5\textwidth}
        \include{compare_lcss_ft2}
        \caption{\texttt{FT-2} configuration}
    \end{subfigure}
    }
    \caption{LCSS measure, simulated FRONT and Maybenot FRONT}
    \label{tab:front-lcss}
\end{figure*}

\begin{figure*}[t!]
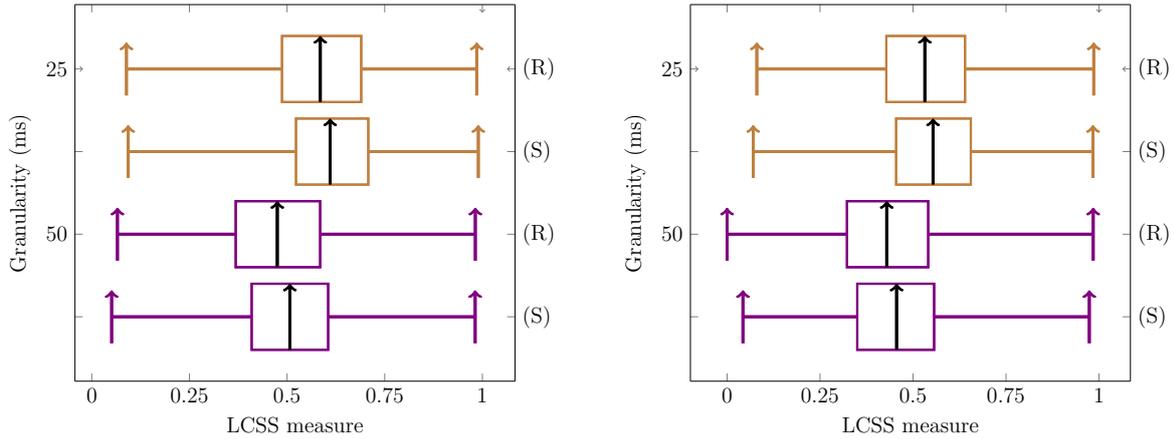

    \centering
    \makebox[\textwidth][c]{
    \begin{subfigure}[b!]{0.5\textwidth}
        \include{compare_lcss_ft1_pipe}
        \caption{\texttt{FT-1} configuration}
    \end{subfigure}
    \hfill
    \begin{subfigure}[b!]{0.5\textwidth}
        \include{compare_lcss_ft2_pipe}
        \caption{\texttt{FT-2} configuration}
    \end{subfigure}
    }
    \caption{LCSS measure, simulated FRONT and Pipelined FRONT}
    \label{tab:front-pipe-lcss}
\end{figure*}

We compared corresponding traces by calculating the Pearson correlation coefficient and a longest common subsequence (LCSS) measure on their matching aggregated time series--once for upload traffic and again for download traffic. We calculated the LCSS measure by dividing the length of the longest common subsequence by the shorter of the lengths of the two aggregated time series being compared. Correlation coefficient results are shown in Figure~\ref{tab:front-correlation} for Maybenot FRONT and Figure~\ref{tab:front-pipe-correlation} for Pipelined FRONT. LCSS results are displayed in Figures~\ref{tab:front-lcss} and~\ref{tab:front-pipe-lcss}.

The correlation coefficient data for both \texttt{FT-1} and \texttt{FT-2} indicates that Maybenot FRONT and Pipelined FRONT padded download traffic similarly to simulated FRONT in most cases. With the \texttt{FT-1} configuration, both defenses have a median correlation of approximately 0.71 at 25 ms granularity and 0.86 at 50 ms granularity. Interquartile range is about [0.56, 0.82] with $I = 25$ and [0.72, 0.93] with $I = 50$ for Maybenot FRONT; it is nearly identical for Pipelined FRONT. Similar results are observed with \texttt{FT-2}.

This indicates a strong correspondence between our implementations and simulated FRONT for at least 75\% of traces. However, a negative correlation is observed for some traces; this is likely due in part to the use of individual states' limit distributions to induce variation of the padding count and window. It is possible for the limit values selected for adjacent \texttt{PADDING} states to differ, which reduces correspondence to the target Rayleigh distribution shape. We also note that each implementation may have selected different values for the padding count and window when defending the same trace, since simulations were run independently; the Appendix provides further discussion.

Upload traffic was not approximated as well as download traffic: with Maybenot \texttt{FT-1} and \texttt{FT-2}, there is a median correlation of about 0.34 when $I = 25$ and 0.48 when $I = 50$; these values drop to 0.25 when $I = 25$ and 0.35 when $I = 50$ with Pipelined FRONT. We attribute this to the higher ratio of padding cells to non-padding cells in upload traffic: in the case of download traffic, there is a higher density of non-padding cells, so any ``misplaced'' padding cells would have a smaller effect on the correlation coefficient.

The minimum median LCSS observed for download traffic is 0.53 when $I = 25$ and 0.43 when $I = 50$ with a narrow interquartile range. Upload traffic has a higher median LCSS in all cases despite lower correlation, which is likely due in part to small quantities of upload traffic resulting in aggregated time series with many low-valued windows. This would occur more often near the end of a trace when less traffic (both padding and non-padding) is sent, suggesting that most of the variation observed in the correlation results is due to differences in padding between implementations for the reasons described above.

\subsubsection{Overhead Measurement}
We continue our evaluation with a comparison of the overhead of simulated FRONT, Maybenot FRONT, and Pipelined FRONT. We consider \textit{bandwidth overhead}, which refers to the total number of padding bytes in a defended trace divided by the total number of non-padding bytes. We further distinguish between \textit{receive} and \textit{send} bandwidth overhead (from the client's perspective), counting cells in only one direction.

\textit{Latency overhead} is another standard metric used in defense evaluations, but FRONT does not delay cells, so no latency overhead is observed in simulated traces. However, we note that ``zero-delay'' defenses \textit{can} cause increased latency in live network deployment scenarios, as described in~\cite{witwer2022padding}. We leave this more robust evaluation to future work and focus here on comparing our implementations to FRONT's expected behavior and providing a preliminary idea of their cost.

\begin{table}[t!]
    \centering
    \begin{tabular}{ |c||c|c|c| }
        \hline
        {Defense} & \multicolumn{3}{c|}{Bandwidth overhead (\%)}\\
        \cline{2-4}
        & \textbf{Send} & \textbf{Receive} & \textbf{Overall} \\
        \hline\hline
        \textbf{Maybenot FT-1} & 597.51 & 41.84 & 78.24 \\
        \hline
        \textbf{Pipelined FT-1} & 613.45 & 43.13 & 80.49 \\
        \hline
        \textbf{Simulated FT-1} & 642.97 & 44.80 & 83.98 \\
        \hline\hline
        \textbf{Maybenot FT-2} & 998.77 & 70.05 & 130.89 \\
        \hline
        \textbf{Pipelined FT-2} & 922.24 & 64.60 & 120.79 \\
        \hline
        \textbf{Simulated FT-2} & 952.91 & 66.24 & 124.32 \\
        \hline
    \end{tabular}
    \caption{FRONT average bandwidth overhead}
    \label{tab:front-overhead}
\end{table}

Mean bandwidth overhead results are presented in Table~\ref{tab:front-overhead}. About 80\% bandwidth overhead was incurred by \texttt{FT-1} and 125\% by \texttt{FT-2}; there was little variation among implementations since the parameters of Maybenot FRONT and Pipelined FRONT were selected to match their bandwidth overhead to that of simulated FRONT.

To accomplish this for Maybenot \texttt{FT-1}, $N$ was decreased from 1700 to 1500. This was necessary because the use of a separate uniform distribution for each \texttt{PADDING} state's limit effectively resulted in a uniform \textit{sum} distribution for padding count, which has a higher expected value. This is also apparent with \texttt{FT-2}, since $N$ was maintained at 2500, and this resulted in 6.57\% greater bandwidth overhead than with simulated FRONT.

However, for Pipelined FRONT, $N$ had to be increased from 1700 to 3000 for \texttt{FT-1} and from 2500 to 4500 for \texttt{FT-2}. We attribute this to the use of pipelines that are based on different padding budgets: there is only a $1 / \psi$ probability of choosing a pipeline that can send $N$ cells, and further reduction of padding count occurs within pipelines.

\subsubsection{Attack Performance}

We evaluated CUMUL~\cite{panchenko2016website}, DF~\cite{sirinam2018deep}, and Tik-Tok~\cite{rahman2019tik} in the closed-world setting against undefended traffic, simulated FRONT, Maybenot FRONT, and Pipelined FRONT. We did this using the scripts provided by Gong et al. for CUMUL~\cite{gong2020zero} and those provided by Rahman et al. for DF and Tik-Tok~\cite{rahman2019tik}. We performed 10-fold cross-validation for all attacks, and we used the model parameters suggested by the attacks' authors, with one exception: the input size of DF and Tik-Tok was changed from 5,000 to 10,000 cells to account for padding. The results are in Table~\ref{tab:front-perf-cw}.

\begin{table}[b!]
    \centering
    \begin{tabular}{ |c||c|c|c| }
        \hline
        {Defense} & \multicolumn{3}{c|}{Accuracy (\%)}\\
        \cline{2-4}
        & \textbf{CUMUL} & \textbf{DF} & \textbf{Tik-Tok} \\
        \hline\hline
        \textbf{Undefended} & 94.66 & 95.89 & 94.00 \\
        \hline\hline
        \textbf{Maybenot FT-1} & 27.68 & 72.11 & 64.00 \\
        \hline
        \textbf{Pipelined FT-1} & 15.72 & 58.00 & 55.89 \\
        \hline
        \textbf{Simulated FT-1} & 12.06 & 48.32 & 49.47 \\
        \hline
        \hline
        \textbf{Maybenot FT-2} & 23.41 & 68.11 & 50.84 \\
        \hline
        \textbf{Pipelined FT-2} & 13.45 & 49.37 & 48.32 \\
        \hline
        \textbf{Simulated FT-2} & 9.12 & 40.95 & 45.79 \\
        \hline
    \end{tabular}
    \caption{FRONT performance in closed-world setting, BigEnough dataset}
    \label{tab:front-perf-cw}
\end{table}

All attacks achieved at least 94\% accuracy on the undefended dataset; these results are similar to those reported by the attacks' authors, but slightly lower values are observed since each class in the BigEnough dataset consists of multiple web pages.

Simulated FRONT decreased CUMUL's accuracy to 12.06\% with the \texttt{FT-1} configuration and 9.12\% with \texttt{FT-2}. It reduced the accuracy of DF and Tik-Tok to 48.32\% and 49.47\%, respectively, with \texttt{FT-1}; and it reduced their accuracy to 40.95\% and 45.79\% with \texttt{FT-2}. These values are used as a benchmark to evaluate our implementations.

Maybenot FRONT was much less effective than simulated FRONT: it reduced Tik-Tok's accuracy to 50.84\% with the \texttt{FT-2} configuration, but it only decreased accuracy to a minimum of 64\% in all other cases. Since there is a high correlation between simulated FRONT and Maybenot FRONT for download traffic but not for upload traffic, this seems to suggest that Maybenot FRONT did not conceal useful features of upload traffic.

However, Pipelined FRONT has a similarly low correlation with simulated FRONT for upload traffic, but it was much more effective than Maybenot FRONT against all attacks: DF was the best attack against it, attaining 58\% accuracy with \texttt{FT-1} and 49.37\% accuracy with \texttt{FT-2}. Thus, we caution that similarity metrics should not be used to conjecture about a defense implementation's protection against attacks; nevertheless, lower values of $I$ might reveal more significant differences between implementations.

We attribute Pipelined FRONT's success to high variation in inter-packet timing, padding count, and padding window, confirming that trace-to-trace randomness can be leveraged to create an effective WF defense. However, simulated FRONT still provided the best protection, highlighting the limits of our approximation approach: FRONT can be implemented effectively with Maybenot, but not precisely.

\section{Implementing RegulaTor}
\subsection{Description}

RegulaTor is based on the observation that Tor traffic consists of occasional ``surges'' of cells sent within a short period of time along with intervening periods of lower cell volume~\cite{holland2022regulator}. Surges are typically present at the beginning of a download, and the amount of traffic decreases exponentially as time elapses. Since most cells are sent in surges, these contain important features that can be leveraged by WF attacks.

In essence, RegulaTor is intended to regularize surges, thereby reducing their uniqueness and, consequently, usefulness as WF attack features. This is accomplished by sending download traffic at a set initial rate which decreases according to a decay function; if the number of queued cells exceeds a threshold, a new surge begins, and traffic is once again sent at the initial rate. Upload traffic is sent at a fraction of the rate of download traffic.

Provisions are also included to reduce overhead and ensure that progress is made. RegulaTor samples a padding count for each download, and it will stop sending padding cells to achieve a constant sending rate after this count has been exceeded, instead delaying non-padding cells to \textit{cap} the sending rate. It will also send any queued upload cells immediately after they have been waiting for a configurable amount of time.

The entire sequence of steps involved in RegulaTor for the relay is:

\begin{enumerate}
  \item Sample a padding count from the discrete uniform distribution \begin{math}[0, N]\end{math}, where \begin{math}N\end{math} is a parameter specifying the padding budget.
  \item Wait until 10 cells are queued, then set the surge start time to the current time.
  \item Set the target rate according to the decay function \begin{math}RD^t\end{math}, where \begin{math}R\end{math} is the initial rate, \begin{math}D\end{math} is the decay parameter, and \begin{math}t\end{math} is time elapsed since the surge start time (seconds).
  \item If the number of queued cells is greater than a threshold parameter, \begin{math}T\end{math}, multiplied by the target rate, reset the surge start time to the current time.
  \item Send a cell if one is queued; otherwise, send a padding cell if the padding count has not yet been exceeded.
  \item Repeat steps 3-5 every time a cell should be sent (determined by the target rate: every \begin{math}rate^{-1}\end{math} seconds) until the download is finished.
\end{enumerate}

The client simply sends cells at a constant fraction of the rate of download traffic: one cell is sent for every \begin{math}U\end{math} cells received. However, if any cells have been queued for more than \begin{math}C\end{math} seconds, they will be sent at once so that downloads continue to make progress.

\subsection {Maybenot RegulaTor}
Two machines were created to approximate RegulaTor, one for clients and one for relays. We refer to these machines collectively as Maybenot RegulaTor.

These machines are based on the \textit{bypass} and \textit{replace} flags of states in Maybenot~\cite{pulls2023maybenot}. If a state with the \textit{bypass} flag set enables blocking, then setting this flag in a padding state allows it to circumvent the blocking and send padding anyway. When \textit{Padding} actions are generated by padding states with the \textit{replace} flag set, the application is allowed to send a queued non-padding cell instead of generating a padding cell.

This combination allows for constant-rate traffic: blocking can be enabled with both flags set, and only padding generated by states with the \textit{bypass} flag set will be allowed through. Such padding can be sent at a constant rate, and by additionally setting the \textit{replace} flag, the application can send non-padding cells instead of padding cells whenever any are queued. When a padding cell is replaced, a \textit{PaddingSent} event is generated as well as a \textit{NonPaddingSent} event. We use this paradigm extensively in Maybenot RegulaTor.

\subsubsection{Relay Machine}

\begin{figure}[b!]
    \centering
    \makebox[\textwidth][c]{
    \begin{tikzpicture}
        \node[state, initial] (q0) {\texttt{START}};
        \node[state, yshift=-96] (q1) {\texttt{BLOCK}};
        \node[state, right of=q1] (q3) {\texttt{BOOT\_0}};
        \node[state, right of=q3] (q4) {\textit{...}};
        \node[state, right of=q4] (q5) {\texttt{BOOT\_8}};
        \node[state, right of=q4, yshift=-96] (q6) {\texttt{SEND\_0}};
        \node[state, left of=q6] (q7) {\textit{...}};
        \node[state, left of=q7] (q8) {\texttt{SEND\_K}};
        \node[state, accepting, left of=q8] (q9) {\texttt{StateEnd}};
        \draw (q0) edge[left, text width=3cm, align=center] node{\scriptsize{\textit{NonPaddingSent\\100\%}}} (q1)
              (q1) edge[above, text width=3cm, align=center] node{\scriptsize{\textit{BlockingBegin\\100\%}}} (q3)
              (q3) edge[loop above, text width=3cm, align=center] node{\scriptsize{\textit{PaddingSent\\100\%}}} (q3)
              (q3) edge[above, text width=3cm, align=center] node{\scriptsize{\textit{NonPaddingSent\\100\%}}} (q4)
              (q4) edge[above, text width=3cm, align=center] node{\scriptsize{\textit{NonPaddingSent\\100\%}}} (q5)
              (q5) edge[loop above, text width=3cm, align=center] node{\scriptsize{\textit{PaddingSent\\100\%}}} (q5)
              (q5) edge[right, text width=3cm, align=center] node{\scriptsize{\textit{NonPaddingSent\\100\%}}} (q6)
              (q6) edge[above, bend right, text width=3cm, align=center] node{\scriptsize{\textit{LimitReached\\100\%}}} (q7)
              (q7) edge[below, bend right, color=blue] node{} (q6)
              (q7) edge[above, bend right, text width=3cm, align=center] node{\scriptsize{\textit{LimitReached\\100\%}}} (q8)
              (q8) edge[below, bend right, color=blue, text width=3cm, align=center] node{\scriptsize{\textit{NonPaddingSent \\\begin{math}2 / (T * rate)\end{math}}}} (q6)
              (q6) edge[loop below, text width=3cm, align=center] node{\scriptsize{\textit{PaddingSent\\100\%}}} (q6)
              (q8) edge[loop below, text width=3cm, align=center] node{\scriptsize{\textit{PaddingSent\\100\%}}} (q8)
              (q8) edge[above, text width=3cm, align=center] node{\scriptsize{\textit{LimitReached\\100\%}}} (q9);
    \end{tikzpicture}
    }
    \caption{Maybenot RegulaTor relay machine with \begin{math}K\end{math} \texttt{SEND} states}
    \label{fig:regulator_relay_imp}
\end{figure}
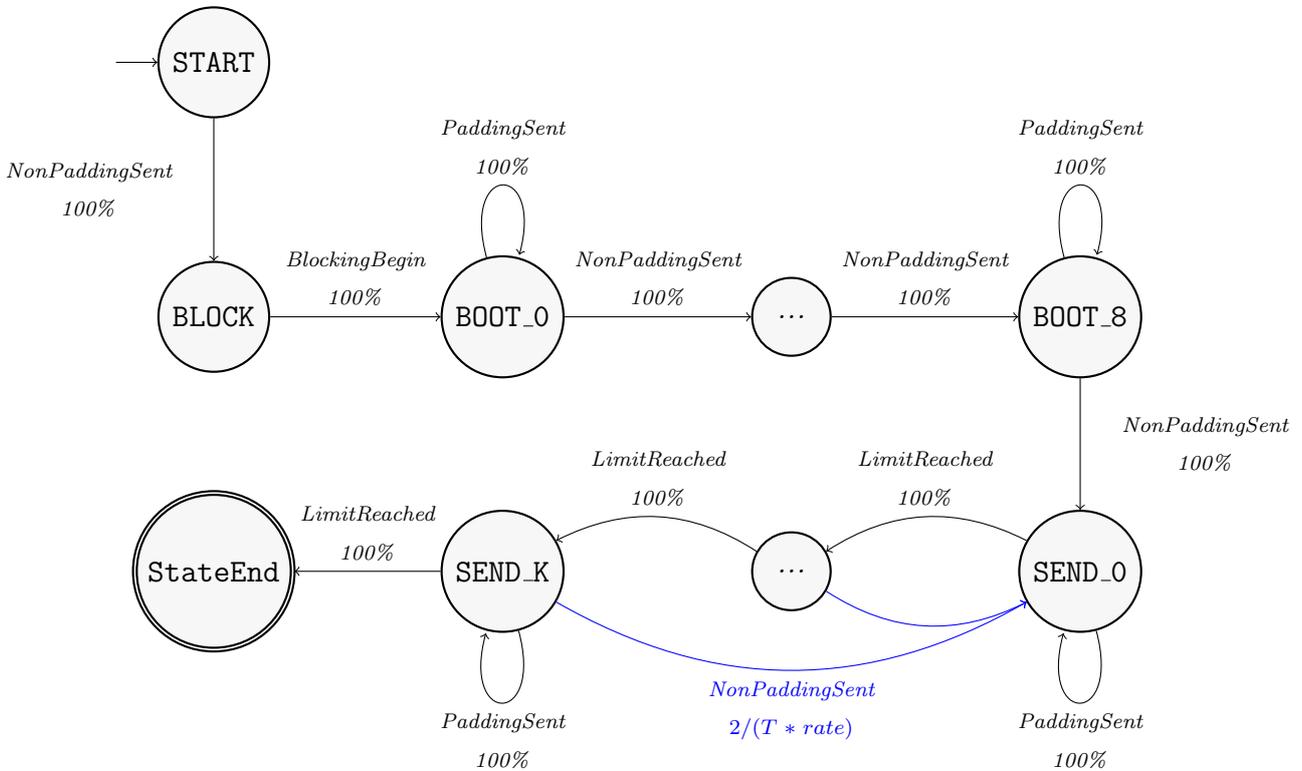

The relay machine can be seen as proceeding through three distinct stages: (1) infinite blocking is enabled with the \textit{bypass} and \textit{replace} flags set; (2) until 10 cells have been sent, a constant traffic rate of 10 cells/second is maintained; and (3) constant-rate \texttt{SEND} states are used to approximate the sending rate imposed by RegulaTor's decay function. The design of this machine is depicted in Figure~\ref{fig:regulator_relay_imp}.

When the first \textit{NonPaddingSent} event is triggered, the machine transitions to the \texttt{BLOCK} state, which enables infinite blocking with the \textit{bypass} and \textit{replace} flags set; this allows for constant traffic rates to be set later, as described previously. Once the application has carried out the blocking action, a \textit{BlockingBegin} event will be triggered, causing the machine to transition to the \texttt{BOOT\_0} state.

Each \texttt{BOOT} state generates a \textit{Padding} action with the \textit{bypass} and \textit{replace} flags set and a 100 ms timeout. When the corresponding \textit{PaddingSent} event is triggered, a self-transition occurs: this results in a constant traffic rate of 10 cells/second. When a \textit{NonPaddingSent} event is triggered, a transition is made to the next \texttt{BOOT} state or, in the case of \texttt{BOOT\_8}, the \texttt{SEND\_0} state. Including the \textit{NonPaddingSent} event that causes a transition to the \texttt{BLOCK} state, then, exactly 10 non-padding cells are sent before the \texttt{SEND\_0} state is reached.
\\
\hfill
\par The \texttt{SEND} states each have the same limit and set a constant traffic rate (timeout) to approximate RegulaTor's decay function. RegulaTor also specifies that if a certain threshold of queued cells is exceeded, a new surge is said to have started and the rate should be increased back to its initial value. Since no queue-related events are present in Maybenot, we implement this behavior probabilistically with a small chance of transitioning back to \texttt{SEND\_0} when a \textit{NonPaddingSent} event is triggered.

Our implementation also excludes the $N$ parameter of RegulaTor, allowing machines to send an unlimited amount of padding during a download. Although Maybenot has features to set padding limits, these will prevent a machine from generating any actions~\cite{pulls2023maybenot}; there is no way to specify that the sending rate should be capped.

\subsubsection{Client Machine}

The client machine sends one cell for every \begin{math}U\end{math} cells received. It consists of a configurable number of \texttt{COUNT} states arranged in sequence, which enable infinite blocking with the \textit{bypass} and \textit{replace} flags set, transitioning to the next state when a \textit{PaddingRecv} or \textit{NonPaddingRecv} event is triggered; and a single \texttt{SEND} state, which generates a \textit{Padding} action with no timeout and the \textit{bypass} and \textit{replace} flags set, transitioning to the first \texttt{COUNT} state when a \textit{PaddingSent} event is triggered.

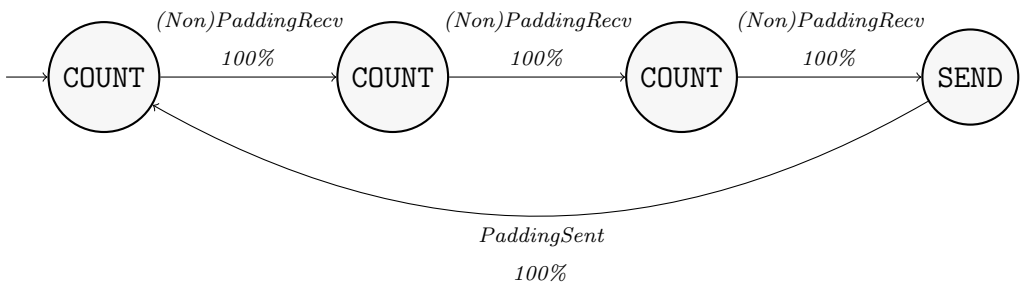
\begin{figure}[b!]
    \centering
    \begin{tikzpicture}
        \node[state, initial] (q1) {\texttt{COUNT}};
        \node[state, right of=q1] (q2) {\texttt{COUNT}};
        \node[state, right of=q2] (q3) {\texttt{COUNT}};
        \node[state, right of=q3] (q4) {\texttt{SEND}};
        \draw (q1) edge[above, text width=3cm, align=center] node{\scriptsize{\textit{(Non)PaddingRecv\\100\%}}} (q2)
              (q2) edge[above, text width=3cm, align=center] node{\scriptsize{\textit{(Non)PaddingRecv\\100\%}}} (q3)
              (q3) edge[above, text width=3cm, align=center] node{\scriptsize{\textit{(Non)PaddingRecv\\100\%}}} (q4)
              (q4) edge[below, bend left, text width=3cm, align=center] node{\scriptsize{\textit{PaddingSent\\100\%}}} (q1);
    \end{tikzpicture}
    \caption{Maybenot RegulaTor client machine with \begin{math}U\end{math} = 3}
    \label{fig:regulator_client_imp}
\end{figure}

If \begin{math}U\end{math} is a whole number, this machine consists of \begin{math}U\end{math} \texttt{COUNT} states that each have a 100\% probability of transitioning to either the next \texttt{COUNT} state or, in the case of the last \texttt{COUNT} state, the \texttt{SEND} state when a \textit{PaddingRecv} or \textit{NonPaddingRecv} event is triggered. Thus, exactly one cell is sent for every \begin{math}U\end{math} cells received; this is the case in Figure~\ref{fig:regulator_client_imp}.

If \begin{math}U\end{math} is \textit{not} a whole number, which is acceptable in RegulaTor, there are \begin{math}\lfloor U \rfloor\end{math} \texttt{COUNT} states, and the probability of transition from the last \texttt{COUNT} state to the \texttt{SEND} state is set to \begin{math}1 - (U - \lfloor U \rfloor)\end{math}; if this does not occur, a self-transition does. The next cell received will cause an immediate transition to \texttt{SEND}: the limit for the last \texttt{COUNT} state is fixed at 2, and the \textit{LimitReached} event causes a transition to \texttt{SEND} with 100\% probability.

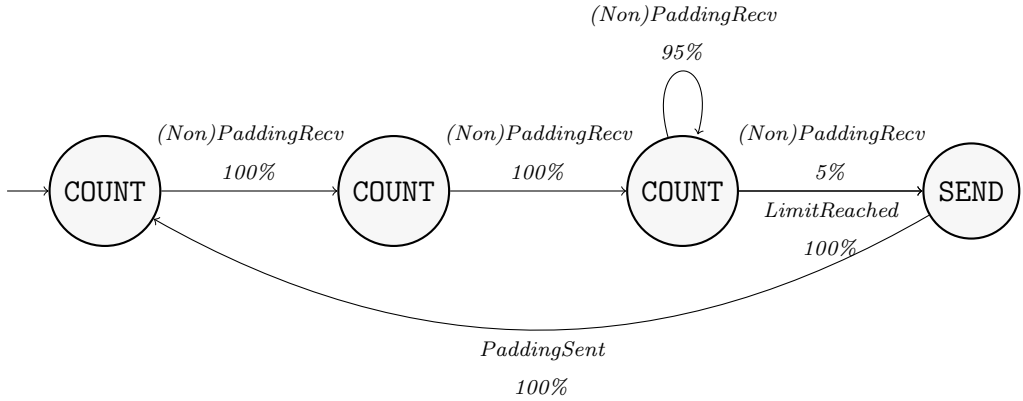
\begin{figure}[t!]
    \centering
    \begin{tikzpicture}
        \node[state, initial] (q1) {\texttt{COUNT}};
        \node[state, right of=q1] (q2) {\texttt{COUNT}};
        \node[state, right of=q2] (q3) {\texttt{COUNT}};
        \node[state, right of=q3] (q4) {\texttt{SEND}};
        \draw (q1) edge[above, text width=3cm, align=center] node{\scriptsize{\textit{(Non)PaddingRecv\\100\%}}} (q2)
              (q2) edge[above, text width=3cm, align=center] node{\scriptsize{\textit{(Non)PaddingRecv\\100\%}}} (q3)
              (q3) edge[above, text width=3cm, align=center] node{\scriptsize{\textit{(Non)PaddingRecv\\5\%}}} (q4)
              (q3) edge[loop above, text width=3cm, align=center] node{\scriptsize{\textit{(Non)PaddingRecv\\95\%}}} (q3)
              (q3) edge[below, text width=3cm, align=center] node{\scriptsize{\textit{LimitReached\\100\%}}} (q4)
              (q4) edge[below, bend left, text width=3cm, align=center] node{\scriptsize{\textit{PaddingSent\\100\%}}} (q1);
    \end{tikzpicture}
    \caption{Maybenot RegulaTor client machine with \begin{math}U\end{math} = 3.95}
    \label{fig:regulator_client_imp2}
\end{figure}

This design is intended to probabilistically approximate the expected behavior of non-integral values of \begin{math}U\end{math}; that is, it should still be the case that one cell is sent for every \begin{math}U\end{math} cells received on average. Thus, if the fractional part of \begin{math}U\end{math} is 0.95, there will be a 5\% chance of reaching \texttt{SEND} after \begin{math}\lfloor U \rfloor\end{math} cells are received and a 95\% chance of reaching \texttt{SEND} after \begin{math}\lceil U \rceil\end{math} cells are received. This situation is depicted in Figure~\ref{fig:regulator_client_imp2}.

While both of these machines effectively mimic the RegulaTor client's behavior, they do not include the \begin{math}C\end{math} parameter, which determines the maximum amount of time that a cell can be queued for before being sent immediately. Thus, a cell might be queued indefinitely, which could result in download progress being slower than with an exact implementation of RegulaTor.

\subsection{Evaluation}

\subsubsection{Experimental Setup}

We used the BigEnough dataset~\cite{mathews2022sok} to evaluate RegulaTor, as detailed for FRONT in Section 3.4. We defended traces with the RegulaTor simulation scripts provided by Holland and Hopper~\cite{holland2022regulator} as well as the Maybenot simulator~\cite{pulls2023maybenot} with the machines described above. Trailing padding cells were removed from the Maybenot-defended datasets before evaluation, as was done for FRONT.

\begin{table}[t!]
    \centering
    \begin{tabular}{ |c||c|c|c|c||c|c||c| }
        \hline
        {Defense} & \multicolumn{7}{c|}{Parameters}\\
        \cline{2-8}
        & \textbf{\begin{math}R\end{math}} & \textbf{\begin{math}D\end{math}} & \textbf{\begin{math}T\end{math}} & \textbf{\begin{math}N\end{math}} & \textbf{\begin{math}U\end{math}} & \textbf{\begin{math}C\end{math}} & \textbf{\begin{math}\omega\end{math}} \\
        \hline\hline
        \textbf{Maybenot RT-Light} & 324 & 0.86 & 3.75 & --- & 4.02 & --- & 20 \\
        \hline
        \textbf{Simulated RT-Light} & 206 & 0.86 & 3.75 & 1650 & 4.02 & 2.08 & --- \\
        \hline\hline
        \textbf{Maybenot RT-Heavy} & 238 & 0.94 & 3.55 & --- & 3.95 & --- & 20 \\
        \hline
        \textbf{Simulated RT-Heavy} & 220 & 0.94 & 3.55 & 2815 & 3.95 & 1.77 & --- \\
        \hline
    \end{tabular}
    \caption{RegulaTor parameters}
    \label{tab:reg-params}
\end{table}

We considered the two configurations of RegulaTor presented by Holland and Hopper: RegulaTor-Light (\texttt{RT-Light}) and RegulaTor-Heavy (\texttt{RT-Heavy}). We derived parameters from the ones used in their paper based on the tuning process they described~\cite{holland2022regulator}. This consisted of multiplying $R$ and $N$ by a ratio between the average number of cells per trace in the BigEnough dataset and the average cells per trace in their dataset. We additionally modified $R$ for Maybenot RegulaTor to match the \textit{latency} overhead of simulated RegulaTor, leaving other parameters unchanged. $\omega$ represents the number of cells per state in Maybenot RegulaTor. All parameters are summarized in Table~\ref{tab:reg-params}.

\subsubsection{Trace Comparison}

\begin{figure*}[t!]
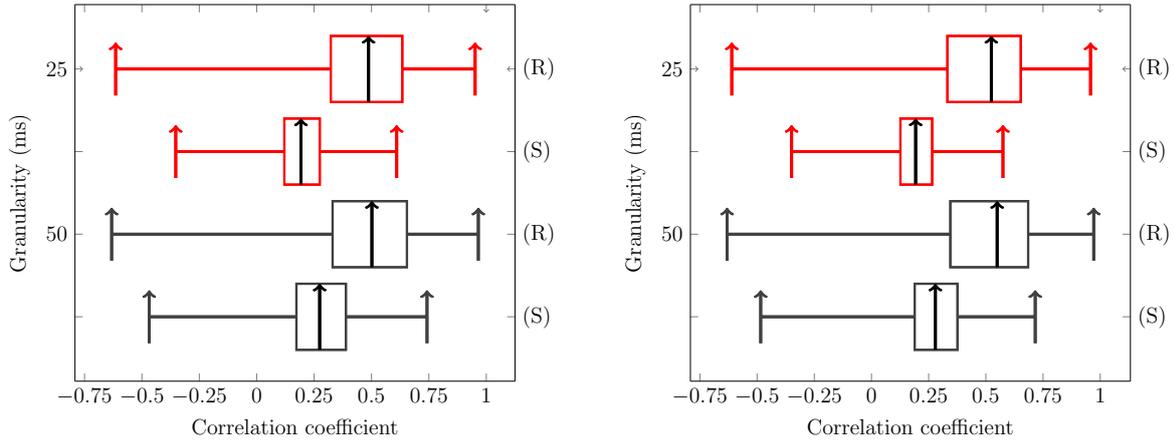

    \centering
    \makebox[\textwidth][c]{
    \begin{subfigure}[b!]{0.5\textwidth}
        \include{compare_corr_reg_light}
        \caption{\texttt{RT-Light} configuration}
    \end{subfigure}
    \hfill
    \begin{subfigure}[b!]{0.5\textwidth}
        \include{compare_corr_reg_heavy}
        \caption{\texttt{RT-Heavy} configuration}
    \end{subfigure}
    }
    \caption{Correlation coefficient, simulated RegulaTor and Maybenot RegulaTor}
    \label{tab:reg-correlation}
\end{figure*}

\begin{figure*}[t!]
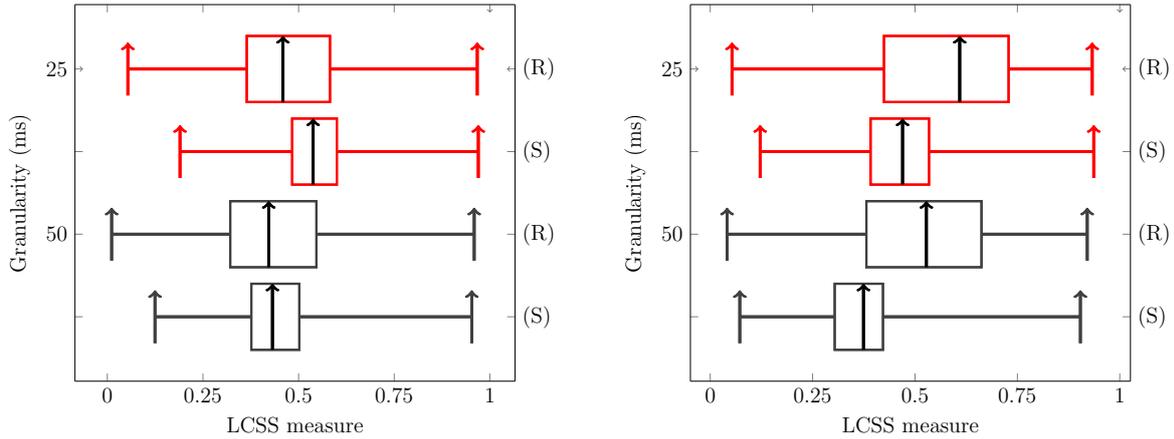

    \centering
    \makebox[\textwidth][c]{
    \begin{subfigure}[b!]{0.5\textwidth}
        \include{compare_lcss_reg_light}
        \caption{\texttt{RT-Light} configuration}
    \end{subfigure}
    \hfill
    \begin{subfigure}[b!]{0.5\textwidth}
        \include{compare_lcss_reg_heavy}
        \caption{\texttt{RT-Heavy} configuration}
    \end{subfigure}
    }
    \caption{LCSS measure, simulated RegulaTor and Maybenot RegulaTor}
    \label{tab:reg-lcss}
\end{figure*}

We generated two aggregated time series for each trace from simulated RegulaTor and Maybenot RegulaTor, using the process described in Section 3.4, with $I = \{25, 50\}$. The correlation results are shown in Figure~\ref{tab:reg-correlation}, and LCSS results are in Figure~\ref{tab:reg-lcss}.

Very similar correlation coefficients are observed with both configurations. There is a moderate correlation for download traffic: with \texttt{RT-Light}, the median is approximately 0.49 with $I = 25$ and 0.50 with $I = 50$. \texttt{RT-Heavy} has a slightly higher median correlation in both cases: it is 0.52 with $I = 25$ and 0.55 with $I = 50$. Interquartile range is $[0.32, 0.63]$ for \texttt{RT-Light} with $I = 25$; similar results are seen for $I = 50$ and \texttt{RT-Heavy}.

Maybenot RegulaTor sent 20 cells per state, allowing for the rate prescribed by the decay function to be matched closely. The moderate correlation observed is likely due to two more significant factors. Maybenot RegulaTor uses a small probability of transitioning to \texttt{SEND\_0} on a \textit{NonPaddingSent} event as a heuristic to mimic RegulaTor's surge restarting behavior, which could result in surges being restarted at different times, greatly decreasing correlation. Also, Maybenot RegulaTor sets a constant rate throughout a download, whereas simulated RegulaTor caps the sending rate after a padding count has been exceeded; this could lead to more divergence towards the end of a trace.

Since upload traffic is simply sent at a constant fraction of the rate of download traffic, the low correlation observed for it is likely due to the same factors and the omission of the $C$ parameter in Maybenot RegulaTor, causing some cells to be sent later.

The median LCSS of \texttt{RT-Light} is about 0.46 with $I = 25$ and 0.42 with $I = 50$; although \texttt{RT-Heavy} has a higher median LCSS in both cases (0.61 with $I = 25$ and 0.53 with $I = 50$), its interquartile range is much greater. This indicates that traces were more similar near the beginning and that much of the observed variation is due to different surge restart times: the probability of restarting a surge in Maybenot RegulaTor decreases as sending rate increases, and sending rate was initially higher with \texttt{RT-Heavy}, resulting in more surge restarts later in a download. Lower LCSS for upload traffic with \texttt{RT-Heavy} also suggests that the $C$ parameter is important, since there was more upload traffic with this configuration and many cells were likely sent later with Maybenot RegulaTor.

\subsubsection{Overhead Measurement}

\begin{table}[b!]
    \centering
    \begin{tabular}{ |c||c|c|c||c| }
        \hline
        {Defense} & \multicolumn{3}{c||}{Bandwidth overhead (\%)} & Latency overhead (\%) \\
        \cline{2-4}
        & \textbf{Send} & \textbf{Receive} & \textbf{Overall} & \\
        \hline\hline
        \textbf{Maybenot RT-Light} & 747.98 & 138.23 & 178.18 & 21.11 \\
        \hline
        \textbf{Simulated RT-Light} & 424.62 & 44.93 & 69.80 & 22.01 \\
        \hline\hline
        \textbf{Maybenot RT-Heavy} & 1091.88 & 151.35 & 212.96 & 15.31 \\
        \hline
        \textbf{Simulated RT-Heavy} & 537.66 & 73.86 & 104.24 & 17.52 \\
        \hline
    \end{tabular}
    \caption{RegulaTor average bandwidth and latency overhead}
    \label{tab:reg-overhead}
\end{table}

We measured both the bandwidth and latency overhead--total time to the last non-padding cell of a trace after being defended compared to original download time--of simulated RegulaTor and Maybenot RegulaTor; mean results are in Table~\ref{tab:reg-overhead}.

Simulated \texttt{RT-Light} incurred 69.80\% bandwidth overhead and 22.01\% latency overhead; \texttt{RT-Heavy} resulted in a higher 104.24\% bandwidth overhead and slightly lower latency overhead (17.52\%), a consequence of its faster sending rate.

With both configurations, Maybenot RegulaTor had comparable latency overhead to simulated RegulaTor, but its bandwidth overhead was much greater: Maybenot \texttt{RT-Light} incurred 178.18\% overhead, a 108.38\% increase over simulated \texttt{RT-Light}; and Maybenot \texttt{RT-Heavy}'s overhead was 212.96\%, which is 108.72\% higher than simulated \texttt{RT-Heavy}.

This is due to the lack of the $N$ parameter in Maybenot RegulaTor: there is no mechanism to limit padding in the relay machine, so a constant traffic rate is set throughout a download. Since surges are restarted probabilistically, it is also likely that this happened more often than necessary. An approximate 108\% increase in bandwidth overhead makes Maybenot RegulaTor too costly for implementation in Tor.

\subsubsection{Attack Performance}

We evaluated CUMUL~\cite{panchenko2016website}, DF~\cite{sirinam2018deep}, and Tik-Tok~\cite{rahman2019tik} in the closed-world setting against simulated RegulaTor and Maybenot RegulaTor using scripts provided by Gong et al. for CUMUL~\cite{gong2020zero} and Rahman et al. for DF and Tik-Tok~\cite{rahman2019tik}. We performed 10-fold cross-validation and changed the input size of DF and Tik-Tok from 5,000 to 10,000 cells to account for padding. The results are in Table~\ref{tab:reg-perf-cw}.

Simulated \texttt{RT-Light} was effective, reducing the accuracy of CUMUL to 5.65\% and DF to 6.42\%, but Tik-Tok attained 22\% accuracy against it. Similarly, simulated \texttt{RT-Heavy} lowered the accuracy of CUMUL to 4.53\% and DF to 5.79\%, and Tik-Tok was the best attack, achieving 15.16\% accuracy.

\begin{table}[t!]
    \centering
    \begin{tabular}{ |c||c|c|c| }
        \hline
        {Defense} & \multicolumn{3}{c|}{Accuracy (\%)}\\
        \cline{2-4}
        & \textbf{CUMUL} & \textbf{DF} & \textbf{Tik-Tok} \\
        \hline\hline
        \textbf{Undefended} & 94.66 & 95.89 & 94.00 \\
        \hline\hline
        \textbf{Maybenot RT-Light} & 6.38 & 6.63 & 9.89 \\
        \hline
        \textbf{Simulated RT-Light} & 5.65 & 6.42 & 22.00 \\
        \hline\hline
        \textbf{Maybenot RT-Heavy} & 6.88 & 8.11 & 10.00 \\
        \hline
        \textbf{Simulated RT-Heavy} & 4.53 & 5.79 & 15.16 \\
        \hline
    \end{tabular}
    \caption{RegulaTor performance in closed-world setting, BigEnough dataset}
    \label{tab:reg-perf-cw}
\end{table}

Maybenot RegulaTor provided better overall protection than simulated RegulaTor. Although CUMUL and DF achieved slightly higher accuracy against it with the \texttt{RT-Light} configuration (6.38\% and 6.63\%, respectively), Tik-Tok's accuracy was only 9.89\%. Similarly, CUMUL and DF were slightly more effective against Maybenot \texttt{RT-Heavy}, but Tik-Tok had lower accuracy than it did against simulated \texttt{RT-Heavy}.

This is likely due to the same factors that increased Maybenot RegulaTor's bandwidth overhead: there was no padding limit, so traffic was sent at a constant rate throughout each download; and surges were restarted probabilistically rather than deterministically, so precise information about the number of queued packets was not leaked. Nevertheless, Maybenot RegulaTor's bandwidth overhead would need to be decreased for it to be a viable candidate for implementation in Tor.

\newpage
\section{Implementing Surakav}
\subsection{Description}

Surakav is intended to achieve the benefits of regularization defenses while decreasing the overhead they often incur~\cite{gong2022surakav}. It works by generating \textit{reference traces}, which appear similar to real cell traces, and using them to shape the traffic pattern of each download. It consists of two components: a \textit{generator}, which uses a GAN to generate reference traces; and a \textit{regulator}, which uses these reference traces to shape traffic.

Both real and reference traces are viewed as \textit{burst sequences}, in which a single burst is an uninterrupted sequence of cells sent in the same direction. An entire burst sequence, then, consists of alternating outgoing/incoming bursts (from the client's perspective), which are characterized by their size, or number of cells.

The defense operates in rounds: during each round, two bursts (outgoing and incoming) are taken from a reference trace. After waiting for a sampled delay, the client sends a burst of real traffic based on the size of the outgoing reference burst. This is followed by a message to the relay to inform it of the size of the incoming reference burst, which it uses to determine the size of its response burst.

The client and relay determine the size of real bursts based on the number of queued cells, the size of the reference burst, and a tolerance parameter $\delta$. A lower and upper threshold are computed as follows, where $b$ is the size of the reference burst:

\begin{equation}
    \bot = \lfloor (1 - \delta) \cdot b \rfloor
\end{equation}

\begin{equation}
    \top = \lfloor (1 + \delta) \cdot b \rfloor
\end{equation}

\hfill
\\
\noindent If the number of queued cells is less than $\bot$, the real burst size is $\bot$; similarly, if it is greater than $\top$, real burst size is $\top$. Otherwise, real burst size is equal to the number of queued cells. Thus, Surakav essentially replays reference traces, modifying burst sizes to reduce overhead; values of $\delta$ closer to zero provide better protection at greater cost.

A mechanism called ``random response'' is also included to decrease overhead: if there are no queued cells at the relay, it can skip sending a response burst with probability $q$, which is sampled from the range $(0, 1)$ for each download. An additional parameter, $\rho$, defines the maximum time gap between outgoing bursts; it is fixed at 100 ms.

\subsection{Maybenot Surakav}

We found that Maybenot is unable to support an accurate implementation of Surakav, principally due to the coordination required between client and relay. We first present two machines that precisely mimic a given reference trace, which we refer to collectively as Maybenot Surakav. We then explore why this design cannot be extended to include burst adjustment or random response.

\begin{figure}[t!]
    \centering
    \makebox[\textwidth][c]{
    \begin{tikzpicture}
        \node[state, initial] (q0) {\texttt{START}};
        \node[state, right of=q0] (q1) {\texttt{BLOCK}};
        \node[state, right of=q1] (q2) {\texttt{SEND}};
        \node[state, right of=q2] (q3) {\texttt{RECV}};
        \node[state, accepting, right of=q3] (q4) {\texttt{StateEnd}};
        \draw (q0) edge[above, text width=3cm, align=center] node{\scriptsize{\textit{NonPaddingSent\\NonPaddingRecv\\100\%}}} (q1)
              (q1) edge[above, text width=3cm, align=center] node{\scriptsize{\textit{BlockingBegin\\100\%}}} (q2)
              (q2) edge[loop above, text width=3cm, align=center] node{\scriptsize{\textit{(Non)PaddingSent\\100\%}}} (q2)
              (q2) edge[above, text width=3cm, align=center] node{\scriptsize{\textit{LimitReached\\100\%}}} (q3)
              (q3) edge[loop above, text width=3cm, align=center] node{\scriptsize{\textit{(Non)PaddingRecv\\100\%}}} (q3)
              (q3) edge[above, text width=3cm, align=center] node{\scriptsize{\textit{LimitReached\\100\%}}} (q4);
    \end{tikzpicture}
    }
    \caption{Maybenot Surakav client machine with two bursts}
    \label{fig:surakav_imp_client}
\end{figure}
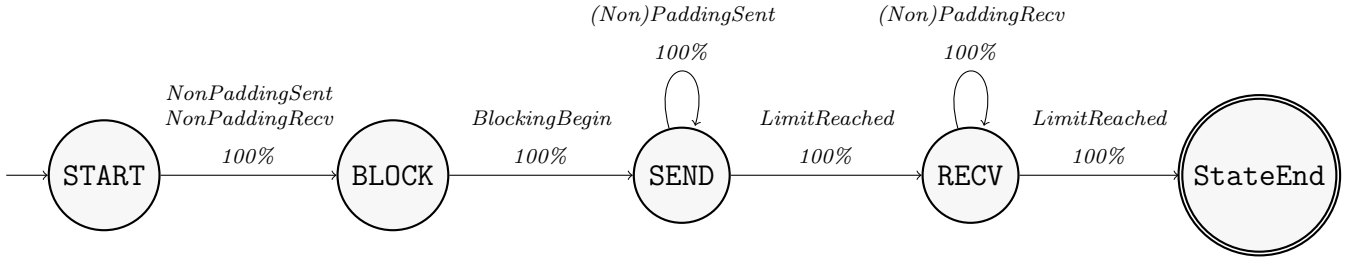

The Maybenot Surakav relay machine is shown in Figure~\ref{fig:surakav_imp_relay}; its corresponding client machine is depicted in Figure~\ref{fig:surakav_imp_client}. These machines consist of a \texttt{START} state; a \texttt{BLOCK} state, which enables infinite blocking with the \textit{bypass} and \textit{replace} flags set; alternating \texttt{SEND} and \texttt{RECV} states; and the pseudo-state \texttt{StateEnd}.

\begin{figure}[b!]
    \centering
    \makebox[\textwidth][c]{
    \begin{tikzpicture}
        \node[state, initial] (q0) {\texttt{START}};
        \node[state, right of=q0] (q1) {\texttt{BLOCK}};
        \node[state, right of=q1] (q2) {\texttt{RECV}};
        \node[state, right of=q2] (q3) {\texttt{SEND}};
        \node[state, accepting, right of=q3] (q4) {\texttt{StateEnd}};
        \draw (q0) edge[above, text width=3cm, align=center] node{\scriptsize{\textit{NonPaddingSent\\NonPaddingRecv\\100\%}}} (q1)
              (q1) edge[above, text width=3cm, align=center] node{\scriptsize{\textit{BlockingBegin\\100\%}}} (q2)
              (q2) edge[loop above, text width=3cm, align=center] node{\scriptsize{\textit{(Non)PaddingRecv\\100\%}}} (q2)
              (q2) edge[above, text width=3cm, align=center] node{\scriptsize{\textit{LimitReached\\100\%}}} (q3)
              (q3) edge[loop above, text width=3cm, align=center] node{\scriptsize{\textit{(Non)PaddingSent\\100\%}}} (q3)
              (q3) edge[above, text width=3cm, align=center] node{\scriptsize{\textit{LimitReached\\100\%}}} (q4);
    \end{tikzpicture}
    }
    \caption{Maybenot Surakav relay machine with two bursts}
    \label{fig:surakav_imp_relay}
\end{figure}
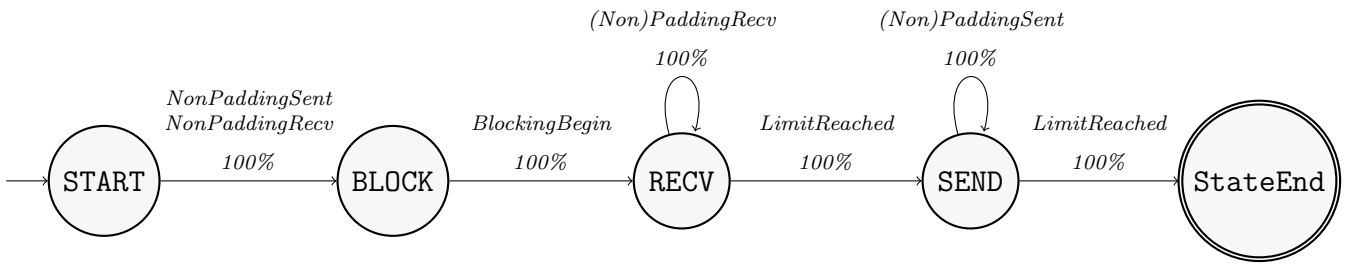

When the first \textit{NonPaddingSent} or \textit{NonPaddingRecv} event is triggered, both machines transition to the \texttt{BLOCK} state, which enables infinite blocking; the \textit{bypass} and \textit{replace} flags are used to allow precise control over the number of outgoing cells. Once blocking is in effect, a \textit{BlockingBegin} event will be triggered, causing the client machine to transition to the \texttt{SEND} state and the relay machine to transition to the \texttt{RECV} state.

In both machines, the \texttt{SEND} state's action is to pad with the \textit{bypass} and \textit{replace} flags. It has a uniform action distribution with parameters $a = 512$ and $b = 512$ to send one cell for each action; a uniform timeout distribution with $a = 5$ and $b = 5$ to delay sending a cell for 5 \textmu s, allowing for a sending rate of 200 cells per second; and a uniform limit distribution which specifies the number of cells to send, based on the size of the corresponding burst from the reference trace.

Each \texttt{RECV} state complements a \texttt{SEND} state; it is configured such that state transitions are synchronized between client and relay machines. The \texttt{RECV} states are set to enable infinite blocking with no timeout and the \textit{bypass} and \textit{replace} flags, which has no effect because this is done immediately by the \texttt{BLOCK} state at the beginning of a download. They have the same limits as their corresponding \texttt{SEND} states.

Though these machines allow for precise replication of a reference trace, they cannot be extended to support burst adjustment due to Maybenot's lack of queue-related events. A further complication is that, even with events for queued cells, queue size information would not be shared by the client and relay: some mechanism would be needed to communicate the size of a burst. Similarly, random response is primarily infeasible because of missing queue information, but it may also require coordination facilities.

We considered the possibility of a heuristic to approximate burst adjustment, as we did with RegulaTor's surge restarting. We created an experimental design that sent up to $\top$ cells per burst and used the number of padding cells sent during a burst as an early stop condition, but the program that generated these machines consistently triggered the out-of-memory killer on our computer with 256 GB of RAM. This is because multiple states were needed per burst, and a single reference trace can contain thousands of these.

Machines must also be sufficiently small in order to be serialized and stored as character strings. Of the 38,000 Maybenot Surakav machine pairs we generated, the majority of serialized machines were about 8.4 MB in size. A machine with more states would require even more storage to be represented, and several machines would need to be stored at a time for future downloads, reducing practicality for ordinary users. This suggests the need for mechanisms to create heuristics that do not require many states.

\subsection{Evaluation}

\subsubsection{Experimental Setup}

For the sake of completeness, we compare simulated Surakav and Maybenot Surakav, despite major differences between implementations. We used the BigEnough dataset~\cite{mathews2022sok} to evaluate Surakav, as described for FRONT in Section 3.4. We defended traces with the Surakav simulation scripts provided by Gong et al.~\cite{gong2022surakav} and the Maybenot simulator~\cite{pulls2023maybenot}. Trailing padding cells were removed from the Maybenot-defended datasets before evaluation, as was done for FRONT and RegulaTor.

We simulated the two configurations of Surakav used by Gong et al.: Surakav-Light ($\delta = 0.6$) and Surakav-Heavy ($\delta = 0.4$). To train the GAN, we used the $CW_{100}$ dataset collected by Rimmer et al. starting in January 2017~\cite{rimmer2018automated}. This dataset consists of 2,500 traces from the homepage of each of the Alexa top 100 websites; we used 1,000 traces from each website. This is similar to the approach taken by Gong et al., who used the $CW_{900}$ dataset, choosing 100 websites at random and using 1,000 traces from each~\cite{gong2022surakav}.

We note that most of the reference traces generated after training with this dataset contained a lot of small bursts. To account for this, we increased the size of the reference trace for each download from 10 to 80 times the size of the undefended trace; however, we set a factor of 480 to defend four particularly unusual traces in the BigEnough dataset. We also modified the Surakav simulation scripts to save the reference traces they generated.

We produced a total of 38,000 Maybenot Surakav machine pairs, one for each configuration and trace in the BigEnough dataset, with the saved reference traces from simulated Surakav. All machines were limited to 8,000 bursts, resulting in many reference traces being truncated, and they did not require any parameters. They were used to create two defended datasets, one with reference traces corresponding to Surakav-Light and another with reference traces from Surakav-Heavy.

\subsubsection{Trace Comparison}

\begin{figure*}[b!]
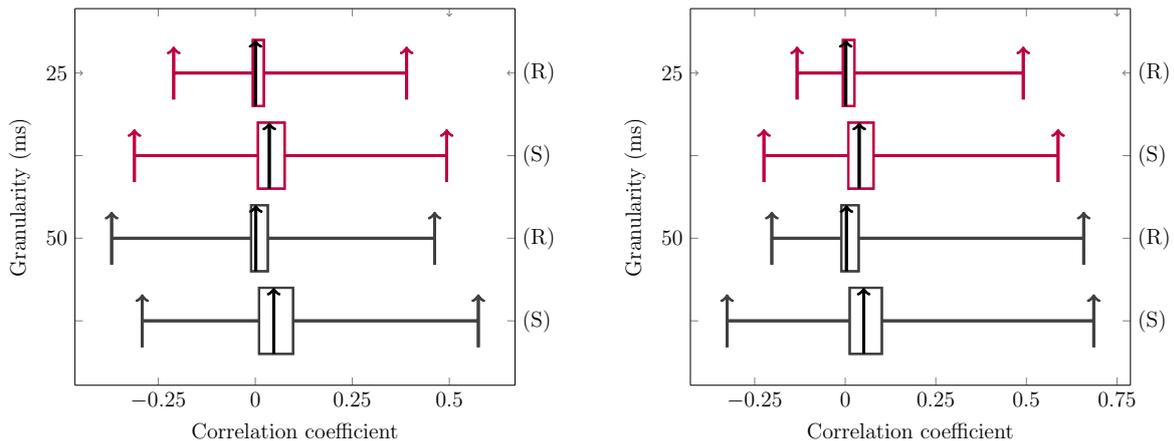

    \centering
    \makebox[\textwidth][c]{
    \begin{subfigure}[b!]{0.5\textwidth}
        \include{compare_corr_sur_light}
        \caption{Surakav-Light configuration}
    \end{subfigure}
    \hfill
    \begin{subfigure}[b!]{0.5\textwidth}
        \include{compare_corr_sur_heavy}
        \caption{Surakav-Heavy configuration}
    \end{subfigure}
    }
    \caption{Correlation coefficient, simulated Surakav and Maybenot Surakav}
    \label{tab:sur-correlation}
\end{figure*}

\begin{figure*}[h!]
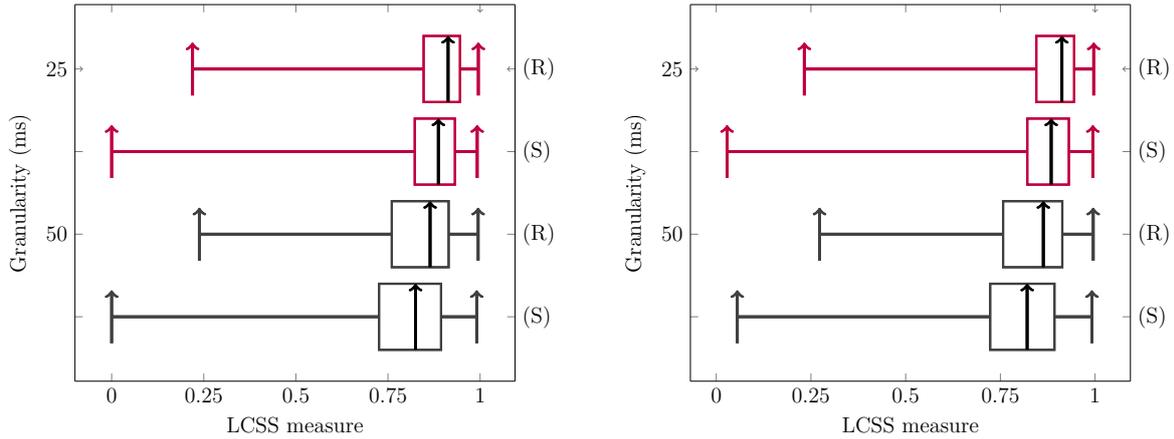

    \centering
    \makebox[\textwidth][c]{
    \begin{subfigure}[b!]{0.5\textwidth}
        \include{compare_lcss_sur_light}
        \caption{Surakav-Light configuration}
    \end{subfigure}
    \hfill
    \begin{subfigure}[b!]{0.5\textwidth}
        \include{compare_lcss_sur_heavy}
        \caption{Surakav-Heavy configuration}
    \end{subfigure}
    }
    \caption{LCSS measure, simulated Surakav and Maybenot Surakav}
    \label{tab:sur-lcss}
\end{figure*}

We generated two aggregated time series for each trace from simulated Surakav and Maybenot Surakav, using the process described in Section 3.4, with $I = \{25, 50\}$. The correlation results are shown in Figure~\ref{tab:sur-correlation}, and LCSS results are in Figure~\ref{tab:sur-lcss}.

As expected, there is a low correlation between simulated Surakav and Maybenot Surakav: the median, 25th percentile, and 75th percentile are nearly zero with both configurations for upload and download traffic. This can be attributed to the lack of burst adjustment and random response in Maybenot Surakav. Upload traffic likely has slightly higher correlation due to its low volume, resulting in less pronounced differences than download traffic.

However, some traces had relatively high correlation: with Surakav-Light, maximum correlation is 0.46 for download traffic and 0.57 for upload traffic when $I = 50$; similar results are observed when $I = 25$. This is likely due to some traces in the BigEnough dataset more closely matching their reference traces. In such cases, simulated Surakav's burst adjustment and random response would not be as pronounced, causing defended traces to be more similar to Maybenot Surakav's. For the same reason, Surakav-Heavy (with $\delta = 0.4$) has a higher maximum correlation: when $I = 50$, it is 0.66 for download traffic and 0.69 for upload traffic.

Interestingly, LCSS is very high for both configurations and values of $I$: with Surakav-Light, median LCSS is 0.89 for upload traffic and 0.91 for download traffic when $I = 25$; it is 0.82 for upload traffic and 0.86 for download traffic when $I = 50$. High median LCSS is also observed with Surakav-Heavy. Maybenot Surakav traces are quite long since more bursts must be sent in total without burst adjustment or random response; thus, we suspect that large portions of simulated Surakav traces were found to be subsequences of their corresponding Maybenot Surakav traces, resulting in a very high LCSS measure.

\subsubsection{Overhead Measurement}

We measured both the bandwidth and latency overhead of simulated Surakav and Maybenot Surakav; mean results are in Table~\ref{tab:sur-overhead}.

\begin{table}[t!]
    \centering
    \begin{tabular}{ |c||c|c|c||c| }
        \hline
        {Defense} & \multicolumn{3}{c||}{Bandwidth overhead (\%)} & Latency overhead (\%) \\
        \cline{2-4}
        & \textbf{Send} & \textbf{Receive} & \textbf{Overall} & \\
        \hline\hline
        \textbf{Maybenot Surakav-Light} & 15601.91 & 226.57 & 1233.81 & 257.66 \\
        \hline
        \textbf{Simulated Surakav-Light} & 531.53 & 65.51 & 96.04 & 31.36 \\
        \hline\hline
        \textbf{Maybenot Surakav-Heavy} & 15514.69 & 225.42 & 1227.02 & 257.14 \\
        \hline
        \textbf{Simulated Surakav-Heavy} & 608.66 & 92.47 & 126.29 & 33.05 \\
        \hline
    \end{tabular}
    \caption{Surakav average bandwidth and latency overhead}
    \label{tab:sur-overhead}
\end{table}

Simulated Surakav-Light incurred 96.04\% bandwidth overhead and 31.35\% latency overhead. With Surakav-Heavy, bandwidth overhead was 125.30\%, and latency overhead was 33.05\%; both of these increases are due to lower tolerance for burst adjustment.

Maybenot Surakav's bandwidth and latency overhead were significantly higher than those of simulated Surakav. With the Surakav-Light configuration, bandwidth overhead was 1233.81\%, and latency overhead was 257.66\%. Similarly, with Surakav-Heavy, bandwidth overhead was 1227.02\%, and latency overhead was 257.14\%.

These results confirm the need for burst adjustment and random response in Surakav: requiring that reference traces be replayed exactly imposes too strict of a traffic pattern, resulting in very high overheads, which make Maybenot Surakav impractical for implementation in Tor in its current state.

\subsubsection{Attack Performance}

We evaluated CUMUL~\cite{panchenko2016website}, DF~\cite{sirinam2018deep}, and Tik-Tok~\cite{rahman2019tik} in the closed-world setting against simulated Surakav and Maybenot Surakav using scripts provided by Gong et al. for
CUMUL~\cite{gong2020zero} and Rahman et al. for DF and Tik-Tok~\cite{rahman2019tik}. We performed 10-fold cross-validation and changed the input size of DF and Tik-Tok from 5,000 to 10,000 cells to account for padding. The results are in Table~\ref{tab:sur-perf-cw}.
\\
\hfill
\\

\begin{table}[h!]
    \centering
    \begin{tabular}{ |c||c|c|c| }
        \hline
        {Defense} & \multicolumn{3}{c|}{Accuracy (\%)}\\
        \cline{2-4}
        & \textbf{CUMUL} & \textbf{DF} & \textbf{Tik-Tok} \\
        \hline\hline
        \textbf{Undefended} & 94.66 & 95.89 & 94.00 \\
        \hline\hline
        \textbf{Maybenot Surakav-Light} & 4.08 & 1.05 & 1.58 \\
        \hline
        \textbf{Simulated Surakav-Light} & 15.86 & 20.32 & 23.37 \\
        \hline\hline
        \textbf{Maybenot Surakav-Heavy} & 4.52 & 1.05 & 2.21 \\
        \hline
        \textbf{Simulated Surakav-Heavy} & 15.41 & 15.47 & 15.47 \\
        \hline
    \end{tabular}
    \caption{Surakav performance in closed-world setting, BigEnough dataset}
    \label{tab:sur-perf-cw}
\end{table}

\par The best attack against simulated Surakav-Light was Tik-Tok, which attained 23.37\% accuracy; CUMUL and DF achieved 15.86\% and 20.35\% accuracy, respectively. Simulated Surakav-Heavy was slightly more effective and provided more consistent protection across all attacks: CUMUL reached 15.41\% accuracy against it, while DF and Tik-Tok both had 15.47\% accuracy.

Maybenot Surakav provided a significant level of protection against all attacks due to its exact mimicking of reference traces. With Maybenot Surakav-Light, CUMUL's accuracy was 4.08\%, and the accuracy of both DF and Tik-Tok was less than 2\%. Similar results are seen with Maybenot Surakav-Heavy, though Tik-Tok's accuracy was slightly higher, at 2.21\%. Regardless, Maybenot Surakav would need to have much lower overhead to be feasible for inclusion in Tor.

\newpage
\section{Discussion}
It is worth considering \textit{why} a framework is better than direct implementation of defenses. The primary appeal is increased flexibility: as observed by Pulls, a framework such as Maybenot would allow for evolving defenses, coordination of multiple defenses, and the use of tailored defenses distributed to clients in real time~\cite{pulls2023maybenot}.

All of these benefits relate to the fact that machines are serialized and represented as character strings, so they can be loaded and used in a plug-and-play fashion. Meanwhile, a direct implementation would require recompilation of part or all of an application's codebase; this is even the case with Tor's existing circuit padding framework~\cite{tor-circpad}. Ideally, the inclusion of a framework would give way to an increased emphasis on defense \textit{design}, and the \textit{integration} of defenses would be simple.

However, providing a framework as the sole or primary mechanism for defense implementation may bind authors of defenses to a specific model, which must be sufficiently expressive and capable of supporting effective defense designs. Even if a framework is flexible in terms of the way defenses are integrated, it will not provide much benefit to users of privacy-enhancing technologies if it cannot be used to implement \textit{effective} defenses, and it may end up in disuse, as has occurred with Tor's circuit padding framework.

\subsection{Suggested Improvements to Maybenot}

Our evaluation has shown that Maybenot \textit{can} be used to approximately implement proposed website fingerprinting defenses. However, it lacks a few features that would significantly improve these implementations.

Based on our experience with RegulaTor, we believe that the ability to monitor queues is critical. If Maybenot included an event for cells being queued, a machine could be created to track the \textit{number} of cells queued, but further support would be needed to track the \textit{time} that cells have been in the queue for. One possibility for this is a timer, which would also be useful for other purposes.

Thus, we suggest the inclusion of two new mechanisms: a \textit{PacketQueued} event, which would allow for the implementation of length-based queue thresholds; and a timer, consisting of an action to start it as well as \textit{TimerStart} and \textit{TimerEnd} events, similar to blocking. This may allow for time-based queue thresholds and could also improve the implementations of a variety of defenses, including Surakav.

However, we consider that a counter would be most beneficial in solving the issues encountered with Surakav, and it could complement the \textit{PacketQueued} event, avoiding the necessity of many states or a separate machine to keep track of queue size. One way of implementing this would be with \textit{CounterIncrement} and \textit{CounterDecrement} actions along with an event triggered when the counter's value is zero after being decremented.

With all of our proposed mechanisms, FRONT could be reimplemented to increment a counter for padding count by a sampled value, avoiding the necessity of multiple pipelines and larger, more complex machines. RegulaTor could also be implemented with the $N$ parameter due to the ability to monitor queues based on length, and the $C$ parameter may be possible to approximate with a timer. Although Surakav probably could not be implemented precisely due to the coordination it requires, the addition of queue capabilities, a timer, and a counter may allow for the development of effective heuristics to approximate burst adjustment and random response.

Another consideration relates to the use of multiple machines in synthesis to enact more sophisticated defenses: we considered such designs for the defenses implemented in this work, but Maybenot has no mechanism for deliberate signaling between machines. The only option that seemed viable was to re-enable blocking that was already in effect, which was also done for ``no-op'' states in the Maybenot RegulaTor client machine and Maybenot Surakav machines. The \textit{BlockingBegin} event could then cause transitions in other machines. However, this would be rather cumbersome and would not work in all situations; internal events and actions for signaling could allow machines to more easily cause state transitions in others upon certain conditions being met.

\subsection{Download Distinction and Deployment Context}

Though the features described above would significantly improve Maybenot's support for proposed defenses, they do not directly address the issue of detecting the end of a download. Maybenot has no events related to download completion; this limitation was circumvented in this work by removing trailing padding cells from defended datasets.

It is important to note that these considerations arise because the defenses we have considered were evaluated on individual traces. Deploying Maybenot at the level of Tor circuits may change their behavior in unexpected ways, since multiple TCP connections can be multiplexed over a single circuit, and users can perform concurrent downloads. Even if Arti could run a separate instance of the framework for each connection, this would not fully solve the problem: traces in evaluation datasets are made up of multiple web requests to retrieve content embedded in fetched HTML pages, and determination of which connections should be grouped together would not be a trivial task.

Since guard and middle relays are not aware of individual TCP connections~\cite{dingledine2004tor}, an event for download completion could only be triggered on the client side, and, in light of the issues described above, this would need to happen on a per-connection basis. Other framework designs do not inherently solve this problem, and further modifications to Tor and any applications that use it would likely be required to do so, which is undesirable. We thus advocate for defenses that are designed to work well at the circuit level, and we believe that all of the defenses we have evaluated could do so with a soft stop condition, though their overhead and protection against attacks may be affected.

The timer we propose could be used to implement a soft stop condition by running a dedicated machine that started a timer when a non-padding cell was sent and in some way signaled other machines to transition upon its expiration. We note that if such a condition were based on cells \textit{received}, complications could arise in situations of congestion or other network-related issues, since a download may not have actually finished. We suggest carefully designed soft stop conditions, which may be defense-specific. The necessity of soft stop conditions further justifies the new features we propose for Maybenot.

\subsection{Alternatives to State Machine Frameworks}

Adding the features we propose would have the effect of generalizing the capabilities of Maybenot. It may be envisioned that features could be continually added to support more defenses, but this would add significant complexity to the framework. Moreover, even if this approach were taken, some defenses could still only be approximated in Maybenot due to the nature of state machine frameworks.

Ultimately, adding additional features to Maybenot would have the effect of assimilating its behavior to that of a high-level extension language. This may in fact be a better design for defense implementation: it could allow for a significant level of flexibility and would not necessarily limit defense designers to any particular model. One example of this approach is WFDefProxy, a framework which allows for use of the Go programming language to implement website fingerprinting defenses on Tor bridges~\cite{gong2021wfdefproxy}. The main drawback of WFDefProxy is that deploying defenses on bridges does not account for the entry point to the Tor network (i.e., the bridge itself) being a potential WF attacker.

WF defenses could also be deployed using Flexible Anonymous Networks (FAN)~\cite{rochet2023flexible}: with this scheme, Arti would be modified to contain \textit{hooks} in certain code locations, and \textit{plugins} could be deployed to interact with desired hooks. This could be done on a per-connection basis, allowing defenses to be encoded as FAN plugins which Tor users could then instruct relays to use. Such a design would also allow defenses to operate at the middle relay, accounting for the threat of a malicious guard; but WFDefProxy and FAN plugins represent only two possible alternatives to a state machine framework.

It is also worth considering that support for some types of defenses or certain features may not be necessary; \textit{effective} defenses are the ultimate goal. Unfortunately, it cannot be said with any certainty \textit{which} features are in fact needed and which are not, because attacks are continually improving, and defenses that were previously thought to be highly effective are no longer useful. Given this, it would be wise to choose an implementation strategy for website fingerprinting defenses that allows for a wide range of capabilities, even if these might not be necessary for effective defenses: it is easier not to use certain features than to later implement ones which turn out to be required.

Thus, we do not recommend Maybenot for Arti in its current state, and we draw no particular conclusions about whether a state machine framework is a good way forward, except that there are some defenses that they will only ever be able to approximate. Adding the features we suggest would enhance Maybenot's capabilities significantly, and further evaluation could provide more insights into whether it is a viable candidate for inclusion in Arti. We also believe that exploring the possibility of an alternative model could be a fruitful direction for future work.

\newpage
\section{Conclusions and Future Work}
We presented approximate implementations of FRONT, RegulaTor, and Surakav in the state machine framework Maybenot. We evaluated these implementations in terms of similarity to the simulated versions of the defenses, overhead, and protection against attacks. This evaluation demonstrates that Maybenot has the potential to support effective defenses, but the addition of certain features would greatly enhance its ability to do so.

We recommend improvements to Maybenot and extensive further evaluation before its inclusion in Arti is considered, but we believe that it could be a good framework for defense implementation if made more expressive. We also recommend consideration of alternative models for defense implementation, since they could support more complex defenses which may be needed as attacks continue to improve.

An immediate avenue for future work would be to attempt to implement more defenses in Maybenot. The defense implementations we presented could also be evaluated with different combinations of parameters, which may provide improvements in terms of overhead and protection against attacks. Evaluation could be carried out with additional datasets and attacks, including attacks performed in the one-page setting to better account for the maximal capabilities of an attacker.

Adding the features we propose to Maybenot and evaluating them with new defense implementations represent an abundance of possibilities for future work. Alternative models for website fingerprinting defenses could also be explored, such as Flexible Anonymous Networks. Ideally, such a model would allow for negotiation of defenses with the middle relay in a Tor circuit due to the possibility of the guard relay being a WF attacker, and it should be capable of supporting proposed defenses.

\section*{Availability}
\addcontentsline{toc}{section}{Availability}

The code for all of the described defense implementations is available on GitHub at \newline \url{https://github.com/ewitwer/maybenot-defenses}.

\newpage
\section*{Acknowledgements}
\addcontentsline{toc}{section}{Acknowledgements}

I would like to thank my advisor, Nick Hopper, for his assistance and helpful feedback over the course of this project as well as his guidance throughout my undergraduate career. He has shaped my experience as a researcher, and I truly appreciate his patience and support for my work.\\

\noindent I would also like to thank James Holland for the useful discussions we have had and reviewing this paper. I've learned a good deal about website fingerprinting from my work with him, and his feedback has been very helpful in my research.\\

\noindent I also extend my gratitude to Kangjie Liu for his review and his helpful feedback and suggestions. I appreciate his insightful perspectives on security that I learned about in his Introduction to Computer Security class.\\

\noindent Special thanks to Tobias Pulls for sharing the code and specifications for the Maybenot framework, providing important insights and feedback throughout this project, and reviewing the paper. Without his help, this project would not have been possible.\\

\noindent Additional thanks to Namitha Binu, Jack Milless, Shana Watters, and Alice Zhang for reviewing a draft of this paper and their helpful suggestions.

\newpage
\printbibliography[heading=bibintoc]

\newpage
\section*{Appendix}
\addcontentsline{toc}{section}{Appendix}

To provide further context for the trace comparisons performed for FRONT, RegulaTor, and Surakav, we also compared datasets defended with the simulated version of each defense on separate occasions. This allowed us to quantify how much of the observed differences was due to selection of different parameters for each trace.

\begin{figure*}[b!]
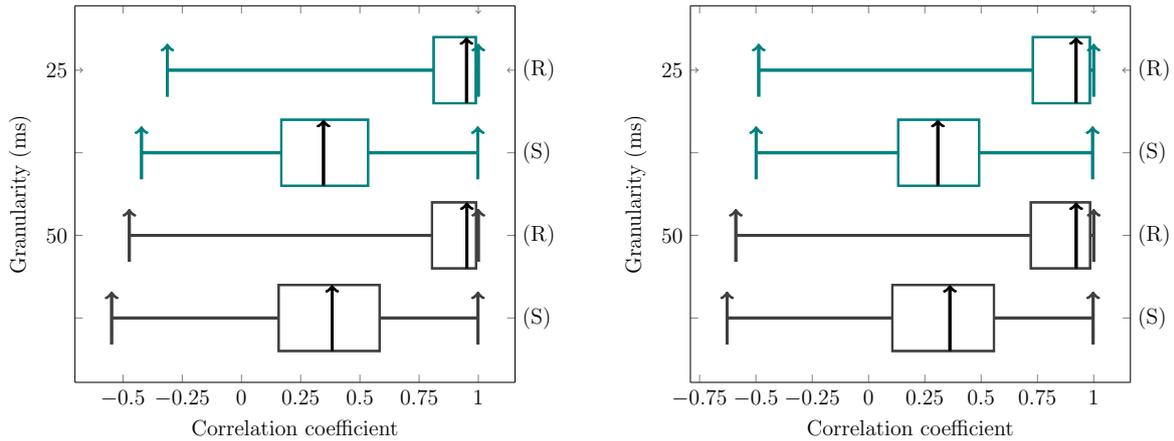

    \centering
    \makebox[\textwidth][c]{
    \begin{subfigure}[b!]{0.5\textwidth}
        \include{compare_corr_ft1ft1}
        \caption{\texttt{FT-1} configuration}
    \end{subfigure}
    \hfill
    \begin{subfigure}[b!]{0.5\textwidth}
        \include{compare_corr_ft2ft2}
        \caption{\texttt{FT-2} configuration}
    \end{subfigure}
    }
    \caption{Correlation coefficient, two runs of simulated FRONT}
    \label{tab:frontfront-correlation}
\end{figure*}

\begin{figure*}[t!]
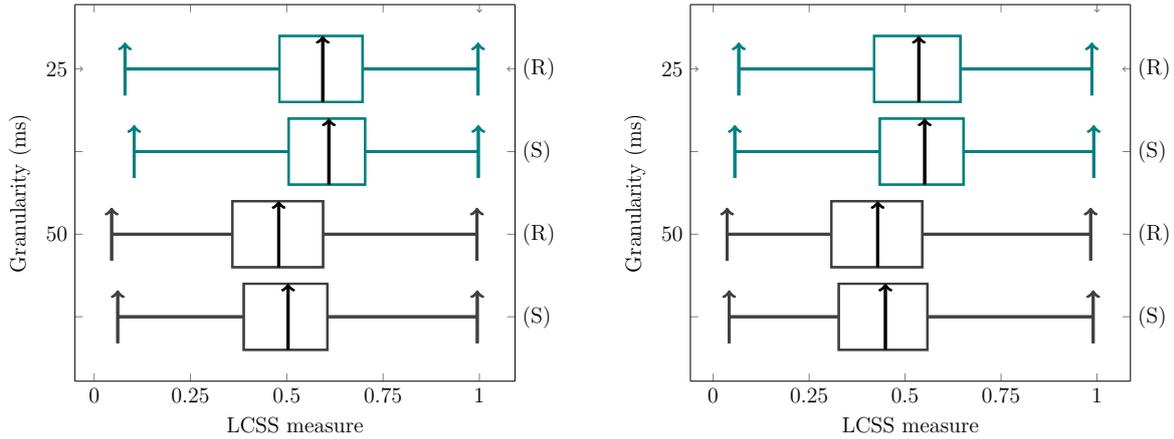

    \centering
    \makebox[\textwidth][c]{
    \begin{subfigure}[b!]{0.5\textwidth}
        \include{compare_lcss_ft1ft1}
        \caption{\texttt{FT-1} configuration}
    \end{subfigure}
    \hfill
    \begin{subfigure}[b!]{0.5\textwidth}
        \include{compare_lcss_ft2ft2}
        \caption{\texttt{FT-2} configuration}
    \end{subfigure}
    }
    \caption{LCSS measure, two runs of simulated FRONT}
    \label{tab:frontfront-lcss}
\end{figure*}

\paragraph{FRONT.} Correlation results for FRONT are shown in Figure~\ref{tab:frontfront-correlation}, and LCSS results are in Figure~\ref{tab:frontfront-lcss}. Median correlation for download traffic is very high: the median is 0.95 with \texttt{FT-1}, and  interquartile range is [0.81, 0.99] when $I = 25$ and [0.80, 0.99] when $I = 50$. Similar results are seen with \texttt{FT-2}. This suggests that the slightly lower correlation with Maybenot FRONT and Pipelined FRONT is due to implementation differences.

There is also a low correlation for upload traffic in all cases: median correlation with \texttt{FT-1} is 0.35 when $I = 25$ and 0.38 when $I = 50$; it is even lower (0.31 when $I = 25$ and 0.36 when $I = 50$) with \texttt{FT-2}. Since low correlation arises from differences only in padding count and window, this supports our conclusion that the low correlation for upload traffic with Maybenot FRONT and Pipelined FRONT can be attributed primarily to a higher ratio of padding to non-padding cells and partly to our implementations.

The minimum median LCSS observed for download traffic is 0.54 when $I = 25$ and 0.43 when $I = 50$. In fact, the results are nearly identical to those of Maybenot FRONT and Pipelined FRONT comparisons; this is likely because padding differences between traces are always present near the beginning of a download, with little variation in the latter portion as few padding cells are sent.

\begin{figure*}[b!]
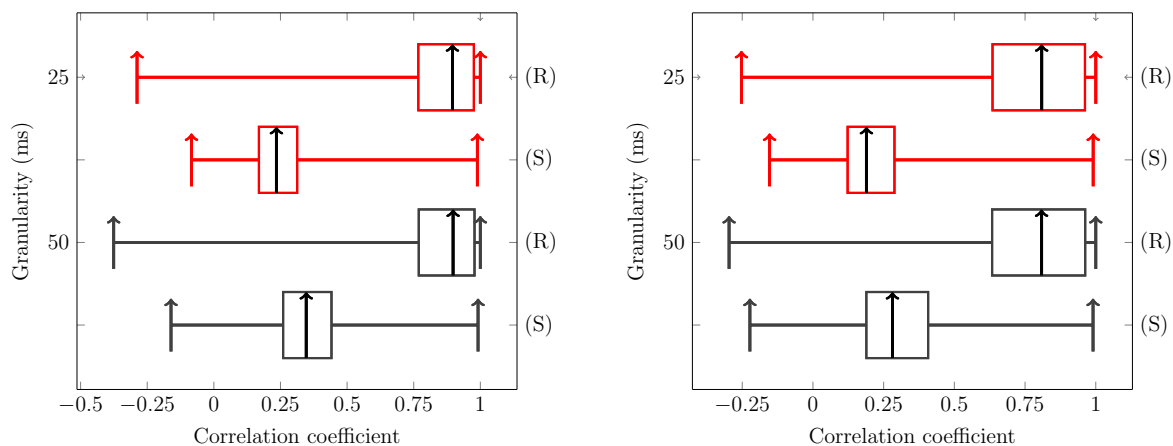

    \centering
    \makebox[\textwidth][c]{
    \begin{subfigure}[b!]{0.5\textwidth}
        \include{compare_corr_regreg_light}
        \caption{\texttt{RT-Light} configuration}
    \end{subfigure}
    \hfill
    \begin{subfigure}[b!]{0.5\textwidth}
        \include{compare_corr_regreg_heavy}
        \caption{\texttt{RT-Heavy} configuration}
    \end{subfigure}
    }
    \caption{Correlation coefficient, two runs of simulated RegulaTor}
    \label{tab:regreg-correlation}
\end{figure*}

\begin{figure*}[b!]
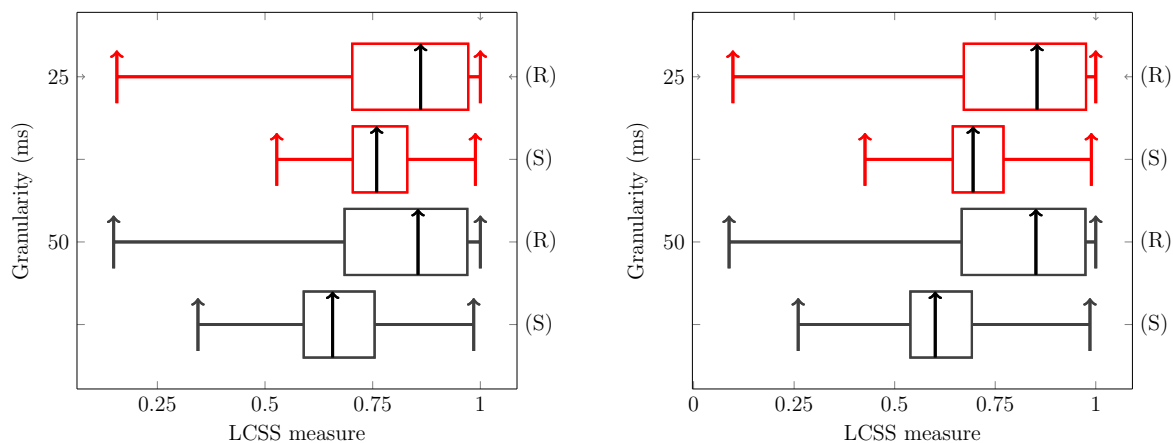

    \centering
    \makebox[\textwidth][c]{
    \begin{subfigure}[b!]{0.5\textwidth}
        \include{compare_lcss_regreg_light}
        \caption{\texttt{RT-Light} configuration}
    \end{subfigure}
    \hfill
    \begin{subfigure}[b!]{0.5\textwidth}
        \include{compare_lcss_regreg_heavy}
        \caption{\texttt{RT-Heavy} configuration}
    \end{subfigure}
    }
    \caption{LCSS measure, two runs of simulated RegulaTor}
    \label{tab:regreg-lcss}
\end{figure*}

\paragraph{RegulaTor.} Correlation results for RegulaTor are shown in Figure~\ref{tab:regreg-correlation}, and LCSS results are in Figure~\ref{tab:regreg-lcss}. As with FRONT, there is a high median correlation for download traffic (0.90 with \texttt{RT-Light} and 0.81 with \texttt{RT-Heavy}, both values of $I$) and narrow interquartile range, but median correlation for upload traffic is low. This is because the only variation between traces defended with simulated RegulaTor arises from the selection of different padding counts, which has a stronger effect on upload traffic due to its lower volume.

This also accounts for the strong median LCSS for both download and upload traffic. However, though upload traffic is typically sent at a constant fraction of the rate of download traffic, its median LCSS is lower: the median LCSS for download traffic is 0.86 with \texttt{RT-Light}, but it is only 0.76 when $I = 25$ and 0.66 when $I = 50$ for upload traffic. This may be due to lower padding counts causing queue sizes to increase at the client. Decreasing the padding count lowers the download traffic rate, which in turn lowers the upload traffic rate. This could result in less queued upload cells being sent near the beginning of a download, and more cells will be queued for longer than $C$ seconds, causing them to be sent immediately. This behavior can change the overall traffic pattern.

\hfill
\\

\begin{figure*}[b!]
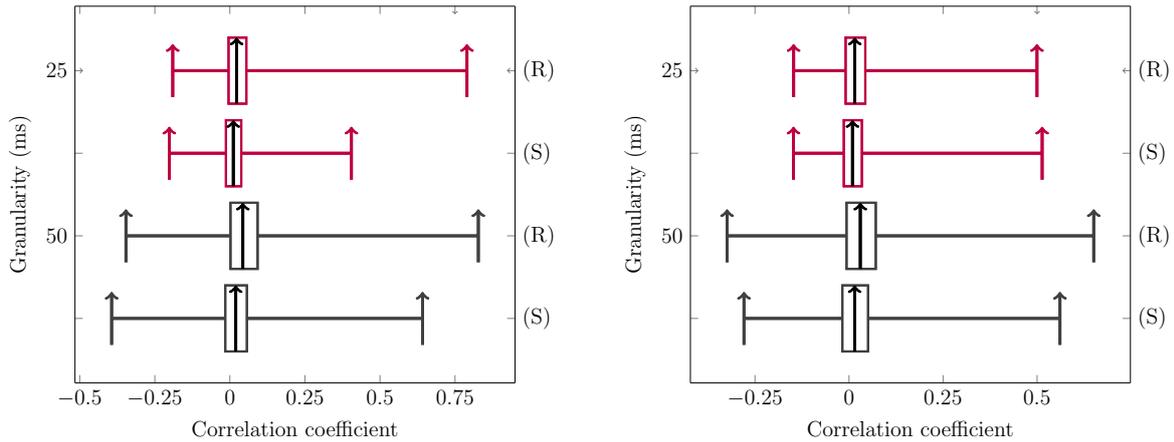

    \centering
    \makebox[\textwidth][c]{
    \begin{subfigure}[b!]{0.5\textwidth}
        \include{compare_corr_sursur_light}
        \caption{Surakav-Light configuration}
    \end{subfigure}
    \hfill
    \begin{subfigure}[b!]{0.5\textwidth}
        \include{compare_corr_sursur_heavy}
        \caption{Surakav-Heavy configuration}
    \end{subfigure}
    }
    \caption{Correlation coefficient, two runs of simulated Surakav}
    \label{tab:sursur-correlation}
\end{figure*}

\begin{figure*}[t!]
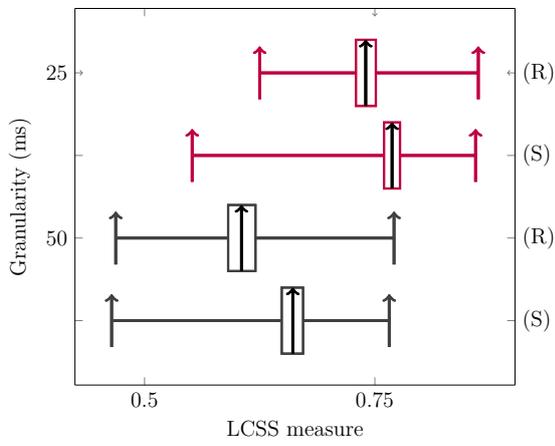
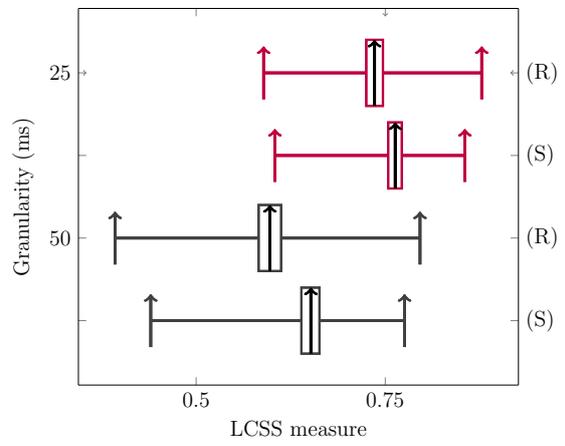

    \centering
    \makebox[\textwidth][c]{
    \begin{subfigure}[b!]{0.5\textwidth}
        \include{compare_lcss_sursur_light}
        \caption{Surakav-Light configuration}
    \end{subfigure}
    \hfill
    \begin{subfigure}[b!]{0.5\textwidth}
        \include{compare_lcss_sursur_heavy}
        \caption{Surakav-Heavy configuration}
    \end{subfigure}
    }
    \caption{LCSS measure, two runs of simulated Surakav}
    \label{tab:sursur-lcss}
\end{figure*}

\paragraph{Surakav.} We reused the same reference traces to compare different runs of simulated Surakav, so any variation should be due only to different random response probability between compared traces. Correlation results for Surakav are displayed in Figure~\ref{tab:sursur-correlation}, and LCSS results are in Figure~\ref{tab:sursur-lcss}.

Surprisingly, median, 25th percentile, and 75th percentile correlation is nearly zero in all cases. This suggests that the random response probability selected for each download can have a significant impact on the resultant defended trace. As a consequence, if Maybenot Surakav were updated to include burst adjustment and random response, it may be expected that the correlation results we report would not change much.

However, LCSS varies markedly from that of simulated Surakav and Maybenot Surakav. With Surakav-Light, the median LCSS for download traffic is 0.74 when $I = 25$ and 0.61 when $I = 50$; median LCSS is 0.77 when $I = 25$ and 0.66 when $I = 50$ for upload traffic. Similar results are seen with Surakav-Heavy. This is likely because corresponding defended traces are of comparable length (as opposed to Maybenot Surakav defended traces), and the decision to skip a burst at the relay can have cascading effects on the remainder of a trace which are more apparent with greater values of $I$.

\end{document}